\documentclass[a4paper,11pt]{article}
\pdfoutput=1 
\usepackage{graphicx}
\usepackage{float}
\usepackage{mciteplus}
\usepackage{slashed}
\usepackage{dirtytalk}
\usepackage{bbold}
\usepackage{braket}
\usepackage{mathtools,amsmath,amssymb,mathbbol}

\usepackage{jhep} 

\usepackage[T1]{fontenc} 

\renewcommand{\[}{\begin{equation}\begin{aligned}}
\renewcommand{\]}{\end{aligned}\end{equation}}
\def\beq{\begin{equation}}
\def\eeq{\end{equation}}

\renewcommand{\texttt}{{}}
\def\bs{\begin{subequations}}
\def\es{\end{subequations}}

\def\Cc{\mathcal{C}}

\def\Ec{\mathcal{E}}
\def\Fc{\mathcal{F}}

\def\Hc{\mathcal{H}}

\def\Kc{\mathcal{K}}
\def\Lc{\mathcal{L}}

\def\Oc{\mathcal{O}}
\def\Pc{\mathcal{P}}

\def\Tc{\mathcal{T}}

\newcommand{\tia}[1]{}

\newcommand{\bea}{\begin{eqnarray}}
\newcommand{\eea}{\end{eqnarray}}
\newcommand{\beas}{\begin{eqnarray*}}
\newcommand{\eeas}{\end{eqnarray*}}
\newcommand{\bal}{\begin{aligned}}
\newcommand{\eal}{\end{aligned}}

\def\({\left(}
\def\){\right)}

\newcommand{\LF}{\left(}
\newcommand{\RF}{\right)}
\newcommand{\LT}{\left[}
\newcommand{\RT}{\right]}
\newcommand{\cpt}{$\Cc\Pc\Tc$}

\newcommand{\pd}{\partial}

\renewcommand{\imath}{\ensuremath{\mathrm{i}}}
\renewcommand{\vec}[1]{\ensuremath{\mathbf{#1}}}

\title{Towards a unitary formulation of quantum field theory in curved spacetime: the case of Schwarzschild black hole}

\author[a]{K. Sravan Kumar}

\author[b]{Jo\~ao Marto}

\affiliation[a~]{Institute of Cosmology \& Gravitation,
	University of Portsmouth,
	Dennis Sciama Building, Burnaby Road,
	Portsmouth, PO1 3FX, United Kingdom}
\affiliation[b~]{
	Departamento de F\'isica, Centro de Matem\'atica e Aplicações (CMA-UBI), Universidade da Beira Interior, Rua Marquês D'Ávila e Bolama, 6201-001 Covilhã, Portugal}

\emailAdd{sravan.kumar@port.ac.uk}
\emailAdd{jmarto@ubi.pt}

\abstract{We argue that the origin of unitarity violation and information loss paradox in our understanding of black holes (BHs) lies in the standard way of doing quantum field theory in curved spacetime (QFTCS), which is heavily biased on intuition borrowed from classical General Relativity. In this paper, with the quantum first approach, we formulate a so-called direct-sum QFT (DQFT) in BH spacetime based on a novel formulation of discrete spacetime transformations in gravity that potentially restores unitarity. By invoking the quantum effects associated with the gravitational backreaction, we show that the Hawking quanta emerging outside of the Schwarzschild radius ($r_S=2GM$) cannot be independent of the quanta that continue to be inside $r_S$. This enables the information to be carried by Hawking quanta, but in the BH DQFT formalism, we do not get any firewalls.  Furthermore, DQFT leads to the BH evaporation involving only pure states. This means the quantum mechanical effects at the BH horizon produce two components of a maximally entangled pure state in geometric superselection sector Hilbert spaces. This construction enables pure states to evolve into pure states, restoring unitarity and observer complementarity.  
Finally, we discuss how our framework leaves important clues for formulating a scattering matrix and probing the nature of quantum gravity.}


\makeatother

\makeatletter
\gdef\@fpheader{}
\makeatother

\begin{document}
	
\maketitle

\section{Introduction}
Ever since Hawking realized black holes (BHs) are unstable quantum mechanically, the world of theoretical physics has taken a murky turn with conceptual conundrums such as unitarity and information loss paradoxes \cite{Hawking:1975vcx,Hawking:1974rv,Hawking:1976ra,Page:1979tc,Page:1993wv,Page:2013dx}.  Even classically, according to general relativity (GR), BHs are the most fascinating objects in the Universe which awe-inspire us to investigate and speculate. The former is arduously difficult, but the latter is a less complicated task because it is based on our natural human intuition/imagination, which emanates from our extreme enthusiasm for understanding  BHs. Our indelible quest for quantum gravity in the last decades has been most often doomed by the interference of our natural imagination and speculations \cite{deBoer:2022zka,Maldacena:2013xja,Maldacena:2018izk,Harlow:2022qsq,Almheiri:2020cfm,Buoninfante:2021ijy}. The main aim of this paper is to go very meticulously through the popularly 'accepted' calculation by Hawking \cite{Hawking:1975vcx} in quantizing a Klein-Gordon (KG) field in the 'near-horizon' spacetime of Schwarzschild black hole (SBH). We embrace the well-known fact that 'time' is a parameter (not an operator) in quantum theory, and one must separate it from the 'time' as we know from GR. In BH physics, we point out that the origin of unitarity violation and information paradox lies in the way notions of time are treated in quantization. To be more precise, in the original Hawking treatment of quantum field theory in curved spacetime  (QFTCS), classical intuitions regarding the arrow of time according to GR are put forward before understanding quantum fields in curved spacetime. However, we note that the direction of exploring nature is from quantum to classical limits rather than the opposite. Thus, we need the quantum first approach, which puts forward the rules of quantum theory before our classical understanding of spacetime \cite{Giddings:2022jda}. It is worth remembering that the arrow of time in quantum physics emerges from the way we impose initial and final conditions on the quantum states \cite{Hartle:2013tm}. Thus, classically, when we see an arrow of time, we must understand it as the thermodynamic arrow of time. Since we carefully go through the conceptual jurisdictions of quantum theory in understanding (quantum) BHs, we request our readers to avoid any preconceived notions fixed in the literature, textbooks, and also popular science articles. 

As it was elaborately discussed in a previous paper \cite{Kumar:2023ctp}, the standard quantum field theory in Minkowski spacetime is very well routed in {the definition of positive energy state with respect to an arrow of time.} 
{We illustrate in this paper that a consistent QFTCS at the horizon of Schwarzchild BH involves two arrows of time, which challenges the foundations of quantum field theory in Minkowski spacetime. For a consistent QFTCS, there are two criteria one must satisfy: one is the unitarity, i.e., an asymptotic far-away observer must see pure states evolving into pure states irrespective of what is falling into the BH, and the second is an infalling local observer at the BH horizon should see a vacuum and no firewalls satisfying the equivalence principle. }
We argue that to achieve a consistent unitary QFTCS, we must step back to identify the discrete spacetime symmetries of the curved manifold, which tells us how to move forward with quantization. In the context of SBH, the question arises as the curvature invariants diverge near the singularity at $r=0$. We stress that one really needs to work with spherical harmonics (with the Kruskal-Szekeres metric) to understand quantum fields in BH backgrounds. Hawking original paper was about quantizing the KG field in Schwarzschild spacetime by making heavy use of spherical harmonics under the assumption that the background nature of spacetime remains almost the same under perturbations. 
Moreover, the effect of curved spacetime occurs for the modes whose wavelengths are comparable to that of the Schwarzschild horizon radius ($r_S=2GM$). If, at these basic levels, we see a unitarity violation problem, which is the fact that there will be an entangled mixed state that can be lost beyond the horizon, resulting in the observer's universe with mixed states. This is called the situation of pure states evolving into mixed states (unitarity violation).  
It is a legitimate quest to save unitarity in the low-energy effective theory, which is just GR and quantum mechanics (together, they should lead to a consistent QFTCS). Thus, to clarify our agenda in the rest of the paper,  we restrict ourselves to refining some steps of Hawking's original paper of 1975 \cite{Hawking:1975vcx} and forget, for the moment, about all developments and the discussion of various possibilities proposed in the name of quantum gravity approaches. We request the respected reader to stay with us in our hunt for the resolution of paradoxes in BH (quantum) physics at the fundamental level within perturbative QFTCS. Our assumption is that the solution to the unitarity violation must be found in the initial approach of Hawking calculation\footnote{Of course, needless to say, we totally acknowledge Hawking's amazing efforts of opening a door into the world of (quantum) BHs back in that time. Hawking has shown nature's most perplexing objects to be unstable quantum mechanically. Still,  paradoxes emerged later on. Therefore, it is essential to ask critical questions.} We thus offer a new understanding of the nature of Hawking radiation beyond those notions widely discussed in the literature \cite{Mathur:2009hf,Mathur:2011uj,Calmet:2022swf}. 

The paper is organized as follows. In Sec.~\ref{sec:onetotwo}, we present how we end up with equally possible descriptions of one physical world with two arrows of time in the context of Schwarzschild BH spacetime. We then present the viewpoint of our study regarding quantum physics at the SBH horizon, which is physically different from existing formulations in the literature. In Sec.~\ref{sec:DQM}, we describe the foundational construction of direct-sum quantum mechanics where a single quantum state can be described with two arrows of time at parity conjugate points.
Sec.~\ref{sec:revMin}, we review the direct-sum QFT (DQFT) in Minkowski spacetime established in \cite{Kumar:2023ctp}. DQFT of Minkowski spacetime gives us the most valuable inputs to understand BHs, basically static curved spacetimes asymptotically Minkowski. The most important aspect of DQFT is the explicit construction of quantum fields as two components which are parity ($\Pc$) and time reversal ($\Tc$) mirror images of each other as per $\Pc\Tc$ symmetry of Minknowski spacetime.
DQFT presents a (quantum) picture of a conformal diagram that exploits further our understanding of {time} as a parameter in quantum theory. It lays a new foundation for understanding QFTCS. In Sec.~\ref{sec:Schldmet}, we discuss the SBH geometry and the discrete symmetries. Then, we point to the fact that the SBH in spherical coordinates is not the structure of metric if one wants to quantize a field in the background because of the coordinate singularity at the horizon $r_S=2GM$. Thus, it is necessary to choose a good set of coordinates, such as Kruskal-Szekers (KS). However, the most crucial observation is that the KS coordinates replace the horizons with discrete transformations. 
With the status of time being a parameter in quantum theory, the discrete spacetime reflections need to be paid special attention here. We shall also comment on the assumptions made by Hawking before quantizing the fields in SBH spacetime and emphasize the need for careful revision. In Sec.~\ref{sec:DQFTBH}, we apply the DQFT formulation for the SBH in quantizing a KG field where we see an emergence of QFTCS that respects the discrete symmetries of SBH spacetime and gives us an entirely new understanding of Hawking radiation. We clearly show here that, quantum mechanically, the SBH in KS coordinates presents us a picture of SBH spacetime where there cannot be a popularly assumed quantum entanglement between the interior and exterior Hawking quanta. This is because the concept of time and the interior and exterior of SBH are not the same; one cannot use the same quantum theory rules everywhere. Moreover, the SBH metric is known to be static for the exterior $r\gtrsim 2GM$ and physically analogous to Kantowski-Sachs metric in the interior  $r\lesssim 2GM$ as found in \cite{Doran:2006dq}. 
We witness a new structure of Fock space for SBH QFTCS, which separates (by geometric superselection rule\footnote{According to this rule Hilbert spaces are defined corresponding to target spatial region with an arrow of time that defines a positive energy state in that region. Such Hilbert spaces are called geometric superselection sectors. When we write a Hilbert space as a direct-sum of geometric superselection sectors, we form a global quantum state as direct-sum of components that correspond to the sectorial Hilbert spaces. }) the gravitational sub-system (SBH) from the environment, i.e., everything outside the SBH. Then a question arises: Are the interior and exterior Hawking quanta independent of each other? To answer this question, we must take into account how the motion of a particle in the exterior of the SBH can gravitationally affect its interior counterpart. This presents an important issue that requires detailed discussion and a derivation of its quantum mechanical implications within the DQFT framework. Then we develop a new picture for understanding (quantum) BH spacetime and its implications for unitarity and observer complementarity, where we elucidate how there cannot be a Page curve \cite{Page:1993wv,Page:2013dx} and firewalls in the BH DQFT formalism. We add some comments about the evaporating BH DQFT picture and open questions for future investigation. In Sec.~\ref{sec:evapBH}, we demonstrate that the BHs cannot be thought of as ordinary quantum systems, which is against the so-called central dogma \cite{Almheiri:2012rt}, at the same time with BH DQFT we show that the three statements are true i.e., (i) Hawking radiation is described by a pure state, (ii) Low energy effective theory is valid, (iii) No firewalls, a free falling observer sees nothing unusual. Thus, our picture goes against the conjectural claims by AMPS \cite{Almheiri:2012rt}. Finally, in Sec.~\ref{sec:QHQG}, we discuss the implications of our QFTCS to the formulation of S-matrix in BH spacetimes and also for quantum gravity. In Sec.~\ref{sec:Conc}, we conclude with a brief summary of results and some comments on future explorations in (quantum) BH physics. In Appendix.~\ref{sec:Rindler}, we present DQFT in Rindler spacetime \cite{Kumar:2024oxf} and discuss the similarities with the BH case. Throughout the paper, we follow the metric signature $(-+++)$ and set $c=1$.


\section{One physical world with two arrows of time}
\label{sec:onetotwo}
In this section, we discuss how one has to end up with a requirement of quantum theory with two arrows of time when we consider quantum effects in gravity. We review here parts of the detailed investigation carried out in \cite{GKM}. 
The Einstein-Rosen(ER)'s seminal paper of 1935 \cite{Einstein:1935tc} is the first work in history that concerned quantum effects in GR. ER attempted to construct quantum theory near the SBH horizon. To do this, one must remove the coordinate singularity in the 
the SBH metric 
\begin{equation}
	ds^2 = -\LF 1-\frac{2GM}{r} \RF dt^2 + \frac{1}{\LF 1-\frac{2GM}{r} \RF}dr^2 + r^2d\Omega^2\,,
	\label{SBHmet}
\end{equation}
at $r=2GM$, which can be done via the following coordinate transformations, \cite{Griffiths:2009dfa}
\begin{equation}
	\begin{aligned}
ds^2 = -\frac{2GM}{r} e^{-\frac{r}{2GM}}dU dV+ r^2d\Omega^2\,,
\label{KSmet}
\end{aligned}
\end{equation}
where $d\Omega^2 =  d\theta^2+\sin^2\theta d\varphi^2$ is the line element of $S_2$ sphere and 
\begin{equation}
	UV = {16G^2M^2} e^{\frac{r}{2GM}}\LF 1-\frac{r}{2GM}\RF,\quad 
	r= r_S +r_S\, W\LT -\frac{UV}{4er_S^2} \RT
\end{equation}
where $r_S=2GM$ is known as Schwarzschild radius and $W$ is the Lambert function.
The metric \eqref{KSmet} can be rewritten as 
\begin{equation}
	ds^2 = \frac{2GM}{r} e^{1-\frac{r}{2GM}} \LF -dT^2 + dX^2 \RF + r^2d\Omega^2\,,
\end{equation}
where 
\begin{equation}
	\begin{aligned}
	T& = \frac{U+V}{2\sqrt{e}},\quad X= \frac{V-U}{2\sqrt{e}},\quad {\rm for}\quad r>2GM \\ 
		T& = \frac{V-U}{2\sqrt{e}},\quad X= \frac{V+U}{2\sqrt{e}},\quad {\rm for}\quad r<2GM
	\end{aligned}
     \label{KScoor} 
\end{equation}
Here $\LF U,\,V \RF$ or $\LF T,\,X\RF$ are called Kruskal-Szekeres (KS) coordinates. An immediate point to notice from \eqref{KScoor} that the meaning of time $T$ is not the same in the interior and exterior. 
 There are two possibilities to define the KS coordinates for the exterior $r>2GM$ (i.e., $UV<0$) geometry of SBH
\begin{equation}
r>2GM \implies \begin{cases}
	U = -4GM e^{-\frac{u}{4GM}} <0,\quad V=4GM e^{\frac{v}{4GM}}>0 \\ 
		U = 4GM e^{-\frac{u}{4GM}} >0,\quad V= -4GM e^{\frac{v}{4GM}}<0
\end{cases}
\label{rg2M}
\end{equation}
where $u=t-r_\ast,\, v=t+r_\ast$ with $r_\ast = r+2GM\ln\big\vert \frac{r}{2GM}-1\big\vert $ being the tortoise coordinate. 
Thus, there are two choices of $U,\,V$ that can describe the exterior $r>2GM$ geometry of SBH, which means the following discrete transformation can describe the same exterior geometry of SBH
\begin{equation}
	U\to -U,\,V\to -V \implies T\to -T
 \label{dissym}
\end{equation}
The same is true for the interior spacetime $r<2GM$ (i.e., $UV>0$)
\begin{equation}
	r<2GM \implies \begin{cases}
		U = 4GM e^{-\frac{u}{4GM}} >0,\quad V= 4GM e^{\frac{v}{4GM}}>0 \\ 
		U = -4GM e^{-\frac{u}{4GM}} <0,\quad V= -4GM e^{\frac{v}{4GM}}<0
	\end{cases}
 \label{rg2M2}
\end{equation}
In the near horizon approximation $r\approx 2GM$ the metric \eqref{KSmet} becomes 
\begin{equation}
ds^2\Big\vert_{r\approx 2GM} \approx -dT^2+dX^2+r_S^2 d\Omega^2\,,
	\label{nearflat}
\end{equation}
which looks like a flat spacetime at a fixed radius $r_S=2GM$. When we aim to understand quantum fields in curved spacetime, we must define a positive energy state that requires a presumption of an arrow of time ($T$), i.e., 
a state with momentum $k=\vert \textbf{k}\vert $ and energy $E$ should evolve as 
\begin{equation}
 i\frac{\pd\vert \Psi_k \rangle }{\pd T} = E\vert \Psi_k\rangle \implies \vert \Psi_k\rangle = e^{-iET}\vert \Psi_k\rangle_0,\quad T: -\infty \to \infty\,.
\end{equation}
As we discussed before, we can describe the region $r>2GM$ with the opposite arrow of time $T: \infty\to -\infty$ ($U\to -U,\, V\to -V$), and then we end up with another physical world with negative energy state. There are two possibilities to deal with this 
\begin{itemize}
	\item Choose an arrow time: say $T: -\infty \to \infty$ and consider the spacetime region as the physical world and assume the alternative spacetime region with the opposite arrow of time $T: \infty \to -\infty$ as the parallel physical world (that is causally separated) or the unphysical world. 
	\item Describe one physical world (i.e., $r>2GM$) with both arrows of time preserving the causality. Einstein-Rosen's 1935 paper suggests this with a conjecture: "A particle in the physical world has to be described by mathematical bridges between two sheets of spacetime".
\end{itemize}
The above two possibilities can also be applied to the interior $r<2GM$, in which case one chooses the region of spacetime with one arrow of time as BH and treats the spacetime region with the opposite arrow of time as White Hole, which is usually treated as unphysical or a physical object somehow opposite of BH that exists somewhere in the Universe. 
 The first approach has an immediate conceptual difficulty: it is not clear where to find another parallel physical world. If we declare the second world as unphysical, then it is not clear why it appears as a possibility from the beginning and the ambiguity of which world we should consider to be physical. This is a major conundrum in this picture as it involves the individual human choice in fixing the arrow of time ($T: -\infty \to \infty$ or $T: \infty \to -\infty$) and, correspondingly, the nature of the physical world. If we make a choice on the arrow of time before quantization, the $T\to -T$ symmetry of SBH spacetime is broken by hand. 
 However, this first approach is what is widely followed in literature, and we will later see that it is what Hawking adopted in his original calculation in 1974. Surely, the unitarity of quantum theory (i.e., pure states evolving into pure states) is lost for an observer in this framework, and attempts have been made in the frameworks of quantum gravity  \cite{Almheiri:2020cfm,Giddings:2022ipt} to restore it. Furthermore, there are other proposed ideas, such as Planck scale BH remnants, which store information about what has formed a BH in the first place, and eventually, one does not have to worry about unitarity and information loss   \cite{Adler:2001vs}. It has also been thought that the two exterior realizations \eqref{rg2M} indicate two causally separated entangled black holes \cite{Maldacena:2013xja}, which emerged from Hawking-Hartle wavefunction computed from Euclidean path integral (complexifying Schwarzschild time) with saddle point approximation \cite{Hartle:1976tp}. Some investigations in recent years include considering soft hair at the BH horizon to restore information \cite{Calmet:2023gbw,Hawking:2016msc}, which are interesting on their own but they do not address the loss of unitarity that inevitably emerges with the standard methods of QFTCS.   
 
Taking the second approach, on the other hand, demands us to rethink and reformulate the standard understanding of quantum theory with two arrows of time, which is what we do in the next section. This is because since SBH is asymptotically Minkowski, if we incorporate both possibilities \eqref{rg2M} in quantum theory, we must reunderstand how one can have QFT formulation in Minkowski spacetime with two arrows of time.
 	Notably, this second approach was also suggested by Schr\"{o}dinger in 1956 \cite{Schrodinger1956,Parikh_2003} in describing (quantum) de Sitter spacetime and 't Hooft in 2016 \cite{tHooft:2016rrl,tHooft:2016qoo} (followed from the earlier work of Sanchez and Whiting \cite{Sanchez:1986qn}) in the name of antipodal identification i.e., parity and time reversal points in spacetime to describe one physical world. We will demonstrate that this second approach also directs us to construct a unitary quantum field theory in curved spacetime. Thus, our framework is this second approach, where we aim to describe one physical world with two arrows of time. It is entirely different from the first one, which either admits a parallel world (or Universe) or omits entirely regions with opposite arrows of time as unphysical.

 
\section{Direct-sum quantum Mechanics}
\label{sec:DQM}
 
Possibilities of describing one physical world with two arrows of time also exist with the Schr\"{o}dinger equation due to the fact that time is a parameter in quantum theory, not an operator like spatial position, which dates back to Wigner's realization of understanding time reversal with anti-unitary operation. 
In general, a positive energy ($\Ec>0$) state in quantum mechanics (QM) is always defined with respect to an arrow of time 
\begin{equation}
	\vert \Psi\rangle_t= e^{-i\Ec t_p}\vert \Psi\rangle_0,\quad t_p: -\infty \to \infty
	\label{posten}
\end{equation}
which emerges from the form of Schr\"{o}dinger equation 
\begin{equation}
	i\frac{\pd \vert \Psi\rangle}{\pd t_p } = \Ec  \vert \Psi\rangle
	\end{equation}
 We can equivalently write the same \eqref{posten} with the opposite arrow of time, which represents again the same positive energy state due to the sign flip of $i\to -i$ 
 \begin{equation}
 	\vert \Psi\rangle_t= e^{i\Ec t_p}\vert \Psi\rangle_0,\quad t_p: \infty \to -\infty
 	\label{posten2}
 \end{equation}
 The state \eqref{posten2} follows from the Schr\"{o}dinger equation 
 \begin{equation}
 	-i\frac{\pd \vert \Psi\rangle}{\pd t_p } = \Ec  \vert \Psi\rangle
 \end{equation}
 Therefore, even in QM, there are two arrows of time that present a state in the physical world (See also \cite{Donoghue:2019ecz,Donoghue:2020mdd}).
 
\begin{figure}
 	\centering
 	\includegraphics[width=0.7\linewidth]{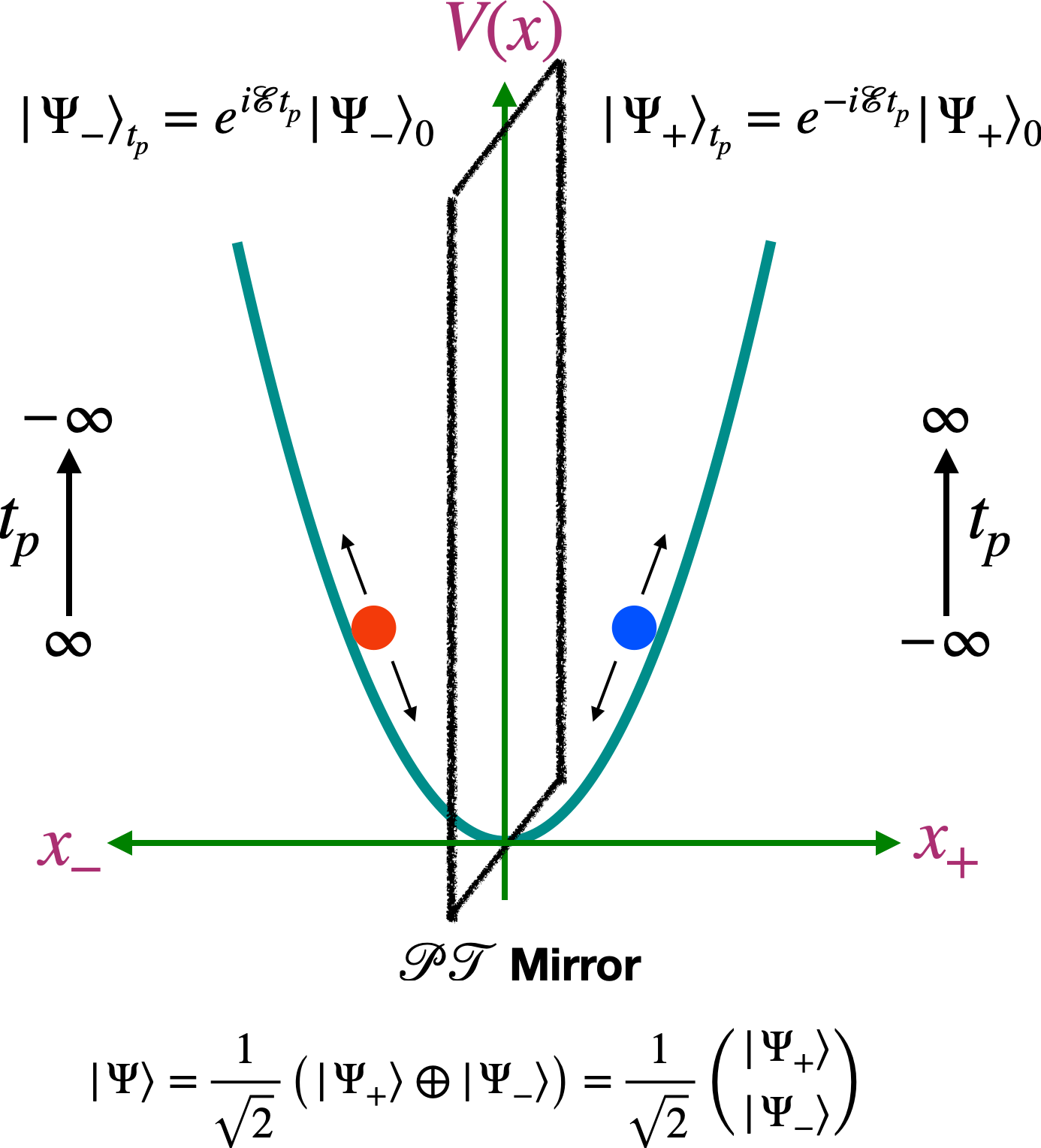}
 	\caption{This picture is a schematic representation of a quantum harmonic oscillator in direct-sum QM. In this picture, we can see the two components of a single quantum state of harmonic oscillators evolving with opposite arrows of time at parity conjugate points of physical space governed by direct-sum Schr\"{o}dinger equation. }
 	\label{fig:ho}
 \end{figure}

Direct-sum QM, which was developed in \cite{Kumar:2023ctp,Gaztanaga:2024vtr} incorporates both arrows of time in the following way by expressing a quantum state as direct-sum of two components\footnote{One must not confuse direct-sum {$ \oplus $}  with the usual summation  {$ +$}. When we take the direct-sum of operators, they take the block diagonal form like \eqref{disum}, and a direct-sum of two state vectors becomes a higher dimensional vector (like \eqref{disumS})\cite{Conway,Harshman,Mazenc:2019ety}}
	\begin{equation}
		\vert \Psi\rangle = \frac{1}{\sqrt{2}} \LF \vert \Psi_+\rangle \oplus \vert \Psi_-\rangle \RF = \frac{1}{\sqrt{2}}\begin{pmatrix}
			\vert \Psi_+\rangle \\ \vert \Psi_-\rangle 
		\end{pmatrix}
		\label{disumS}
	\end{equation}
that evolve with opposite arrows of time at the parity conjugate points of physical space corresponding to geometric superselection sectors\footnote{Superselection sectors are Hilbert spaces whose direct-sum forms the total Hilbert space. According to this, states in the superselection sectors cannot form a superposition with each other \cite{Wick:1952nb,nlab:superselection_theory,Kumar:2023ctp,GKM}. In our context, the superselection sectors are associated geometrically with parity conjugate regions of physical space, which is totally different from what is known in the algebraic QFT \cite{Wick:1952nb, nlab:superselection_theory,Kumar:2023ctp,GKM}. Thus, we name our construction as a quantum theory with geometric superselection sectors, and we will later see it forms the basis for a unitarity quantum field theory in curved spacetime.} of Hilbert space $\Hc = \Hc_+ \oplus \Hc_-$.
The time evolution of the state \eqref{disumS} is governed by what we call the direct-sum Schr\"{o}dinger equation given by
\begin{equation}
i\frac{\pd}{\pd t_p}\begin{pmatrix}
\vert \Psi_+\rangle \\ \vert \Psi_-\rangle 
\end{pmatrix} = \begin{pmatrix}
\hat{H}_+ & 0 \\ 0 & -\hat H_-
\end{pmatrix} \begin{pmatrix}
\vert \Psi_+\rangle \\ \vert \Psi_-\rangle 
\end{pmatrix} 
\end{equation}
where $\hat H= \hat H_+\LF \hat x_+,\, \hat p_+ \RF \oplus  \hat H_-\LF \hat x_-,\, \hat p_-\RF$ is the total (time-independent) Hamiltonian written as direct-sum two components $\hat H_\pm$ which are functions of position and momentum operators $\LF \hat x_\pm,\, \hat p_\pm \RF$. {The eigenvalues of $\hat x_\pm$ are parity conjugate positions $x_+ = x> 0$ (i.e., $x_+\in (0,\,\infty]$) and $x_- = x< 0$ (i.e., $x_-\in [-\infty, 0)$). Note that any translation $x\to x+L$ would result in the corresponding translations $x_\pm \to x_\pm+L$ in the geometric superselection sectors. {Furthermore, a quantum state cannot be moved across geometric superselection sectors by any set of local operations.} } And also, note that the operators of one superselection sector do not act on the states of the other by construction \cite{Conway}, for example 
  \begin{equation}
	\begin{aligned}
		\hat{H}\vert \psi_{+} \rangle & =  \hat{H}_{+}\LF \hat x_+, \hat{p}_+\RF\vert \psi_{+} \rangle \\
		\hat{H}\vert \psi_{-} \rangle & =  \hat{H}_{-}\LF \hat x_-, \hat{p}_-\RF\vert \psi_{-} \rangle
	\end{aligned} 
\end{equation}
The canonical commutation relations in geometric superselection sectors are given by\footnote{Furthermore, note that 
 	\begin{equation}
 		\Big[ \hat x_+,\,\hat x_- \Big]= \Big[ \hat p_+,\,\hat p_- \Big]= \Big[ \hat x_+,\,\hat p_- \Big] = \Big[ \hat p_+,\,\hat x_- \Big] =0\,.
 	\end{equation}
 }
 \begin{equation}
 	\begin{aligned}
 		[\hat{x}_+,\,\hat{p}_+] & = i\hbar\,,\quad \hat{p}_+= -i\hbar \frac{\pd}{\pd x_+}\quad x_+=x> 0 \, \\ 
 		[\hat{x}_-,\,\hat{p}_-] &  = -i \hbar\,,\quad \hat{p}_-= i\hbar \frac{\pd}{\pd x_-}\quad x_-=x< 0 
 	\end{aligned}
 \end{equation}
 The factor of $\sqrt{2}$ in \eqref{disumS} is a normalization to have the total probabilities add up to 1 as 
 \begin{equation}
 	\int_{-\infty}^\infty dx\langle \Psi \vert \Psi \rangle = \int^{\infty}_0 dx_+\frac{\langle \Psi_+ \vert \Psi_+\rangle }{2}+\int^{0}_{-\infty} dx_{-}\frac{\langle \Psi_{-}\vert \Psi_{-}\rangle }{2}  =1\,,
 \end{equation}
 In the direct-sum QM, a single quantum state (positive energy) is represented by direct-sum of two components where one evolves forward at position $x$ according to the arrow of time $t_p: -\infty \to \infty$ and the other evolves backward in time at position $-x$ according to the arrow of time $t_p: \infty \to -\infty$ 
 \begin{equation}
 	\vert \Psi\rangle_{t_p} = \frac{1}{\sqrt{2}} \begin{pmatrix}
 		\vert \Psi_+\rangle_0 e^{-i\Ec t_p} \\ 
 		\vert \Psi_{-}\rangle_0 e^{i\Ec t_p}
 	\end{pmatrix}
 	\label{staespm}
 \end{equation}
 where $\Ec$ is the energy eigenvalue in the case of the time-independent Hamiltonian. The wavefunction is\footnote{See \cite{Gaztanaga:2024vtr,Kumar:2024quv} for the explicit form of the wave function for the quantum harmonic oscillator.}
 \begin{equation}
 	\Psi(x,t_p) = \langle x\vert \psi\rangle = \begin{pmatrix}
 		\langle x_+\vert & \langle x_-\vert 
 	\end{pmatrix} \begin{pmatrix}
 		\vert \Psi_+\rangle \\ \vert  \Psi_-\rangle 
 	\end{pmatrix} =   \frac{1}{2}  \Psi(x_+) e^{-i\Ec t_p} + \frac{1}{2}  \Psi(x_-)e^{i\Ec t_p}
 \end{equation}
 With the above construction, we revisualize the quantum harmonic oscillator through two components of a state, which are "quantum mirror ($\Pc \Tc$)" images of each other (See Fig.~\ref{fig:ho}.)
 Note that any observable in quantum mechanics is now split into two parts 
  \begin{equation}
 	\langle \Psi \vert \hat \Oc \vert \Psi \rangle = 	\langle \Psi \vert \hat \Oc_+\oplus \hat \Oc_- \vert \Psi \rangle =  \frac{1}{2}	\langle \Psi_+ \vert \hat \Oc_+ \vert \Psi_+  \rangle +\frac{1}{2}	\langle \Psi_- \vert \hat  \Oc_- \vert \Psi_- \rangle \,.
 \end{equation}

\section{Direct-sum quantum field theory (DQFT) in Minkowski spacetime} 
\label{sec:revMin}

In this section, we present the construction of direct-sum quantum fields in Minkowski spacetime 
\begin{equation}
    ds^2 = -dt_p^2+d\textbf{x}^2\,.
    \label{minst}
\end{equation}
First of all, it is worth noticing that Minkowski spacetime is symmetric under $\Tc: t_p\to -t_p$ and $\Pc: \textbf{x}\to -\textbf{x}$. 
Let us begin by outlining the implementation of the direct-sum quantization scheme for the Klein-Gordon field: We divide the physical space by parity (say $\textbf{x}$ and $-\textbf{x}$), and when the KG field operator acts on the direct-sum vacuum $\hat{\phi}\vert 0\rangle_T  $ a component of quantum field gives quantum state(s) evolving forward in time at position $\textbf{x}$ while the other component of quantum field leads to state(s) evolving backward in time at position $-\textbf{x}$.

The Klein-Gordon (KG) field operator in the DQFT  formulation is represented as a direct-sum of two components, as written below
\begin{equation}
	\begin{aligned}
		\hat{\phi}\LF x \RF & = \frac{1}{\sqrt{2}}  \hat{\phi}_{+}  \LF t_p,\, \textbf{x} \RF \oplus \frac{1}{\sqrt{2}} \hat{\phi}_{-} \LF -t_p,\,-\textbf{x} \RF \\ 
		& = \frac{1}{\sqrt{2}} \begin{pmatrix}
			\hat{\phi}_{+} & 0 \\ 
			0 & 	\hat{\phi}_{-}
		\end{pmatrix}
	\end{aligned}
	\label{disum}
\end{equation}
where 
\begin{equation}
	\begin{aligned}
		\hat{\phi}_{+}  \LF t_p,\, \textbf{x} \RF &  = 	\int \frac{d^3\vec{k}}{\LF 2\pi\RF ^{3/2}	}	\frac{1}{\sqrt{2 \omega_\vec{k}}} \Bigg[\hat a_{+\,\vec{k}}  e^{ik\cdot x}+\hat a^\dagger_{+\,\textbf{k}} e^{-ik\cdot x} \Bigg] \\ 
		\hat{\phi}_{-}  \LF -t_p,\, -\textbf{x} \RF &  = 	\int \frac{d^3\vec{k}}{\LF 2\pi\RF ^{3/2}	}	\frac{1}{\sqrt{2 \omega_\vec{k}}} \Bigg[\hat a_{-\,\vec{k}}  e^{-ik\cdot x}+\hat a^\dagger_{-\,\textbf{k}} e^{ik\cdot x} \Bigg]\,, 
	\end{aligned}
	\label{fiedDQFTMin}
\end{equation}
where $k\cdot x= -t_p k_0+\textbf{k}\cdot \textbf{x}$. 
Here, components of field $\hat{\phi}_\pm$ in the parity conjugate spatial regions $\textbf{x}> 0$ denoted by "+" and $\textbf{x}< 0$ denoted by "$-$" respectively where they evolve with opposite arrows of time. 
Here the operators $a_+,\, a_{-}$ and  $a_+^\dagger,\, a_{-}^\dagger$ are the annihilation and creation operators, respectively satisfying the canonical commutation relations (in this section we use the units of $\hbar=1$)
\begin{equation}
	\begin{aligned}
		[\hat{a}_{+\,\textbf{k}},\,\hat{a}_{+\,\textbf{k}^\prime}^\dagger] & = 	[\hat{a}_{-\,\textbf{k}},\,\hat{a}_{-\,\textbf{k}^\prime}^\dagger] = \delta^{(3)}\LF \textbf{k}-\textbf{k}^\prime \RF\\
		[\hat{a}_{+\,\textbf{k}},\,\hat{a}_{-\,\textbf{k}^\prime}] &=	[\hat{a}_{+\,\textbf{k}},\,\hat{a}_{-\,\textbf{k}^\prime}^\dagger] = [\hat{a}_{+\,\textbf{k}}^\dagger,\,\hat{a}_{-\,\textbf{k}^\prime}^\dagger]  =0\,.
	\end{aligned}
	\label{comcan}
\end{equation}
and we demand the mixed commutation relations i.e., the second line of \eqref{comcan} such that 
\begin{equation}
	[\hat{\phi}_+\LF x \RF,\, \hat{\phi}_{-}\LF -x^\prime \RF]= 0\,, 
	\label{newcomcond}
\end{equation}
Furthermore, the commutation relations of the field operator components $\hat{\phi}_+$,\, $\hat{\phi}_{-}$ and the corresponding conjugate momenta are defined by
\begin{equation}
	[\hat{	\phi}_+\LF t_p, \textbf{x} \RF,\,\hat{\pi}_+\LF t_p,\,\textbf{x}^\prime \RF] = i \delta\LF \textbf{x}-\textbf{x}^\prime \RF,\quad [\hat{	\phi}_{-}\LF- t_p, -\textbf{x} \RF,\,\hat{\pi}_{-}\LF -t_p,\,-\textbf{x}^\prime \RF] = -i \delta\LF \textbf{x}-\textbf{x}^\prime \RF,
	\label{canDQFT}
\end{equation}
where 
\begin{equation}
	{\pi}_+ \LF t_p,\,\textbf{x} \RF = \frac{\pd\Lc_{\rm KG}}{\pd\LF \pd_{t_p}  \phi_+\RF},\quad  	{\pi}_{-} \LF t_p,\,\textbf{x} \RF = - \frac{\pd\Lc_{\rm KG}}{\pd\LF \pd_{t_p}  \phi_{-}\RF}
\end{equation}
The Fock space (Minkowski) vacuum in the DQFT is the direct-sum of two\footnote{We stress that our construction is not a description of two copies of Minkowski spacetimes (like it is done in \cite{Hartman:2020khs}), but rather it is a description of a quantum field in one Minkowski spacetime with two different vacuums joined by a $\Pc\Tc$ based geometric superselection rule.  } 
\begin{equation}
	\vert 0\rangle_T = \vert 0\rangle_+ \oplus \vert 0\rangle_{-} =  \begin{pmatrix}
		\vert 0\rangle_{+} \\ \vert 0\rangle_{-}\
	\end{pmatrix}\,.
	\label{tFs}
\end{equation}
defined by
\begin{equation}
	\hat{a}_{+\,\textbf{k}}\vert 0\rangle_+ = 0,\quad 	\hat{a}_{-\,\textbf{k}}\vert 0\rangle_{-} = 0\,. 
\end{equation}
The total Fock space is also direct-sum of the two Fock spaces (geometric superselection sectors) as
\begin{equation}
	\Fc_T = \Fc_+ \oplus \Fc_{-}\,,
	\label{disumFock}
\end{equation}
where 
\begin{equation}
	\begin{aligned}
		\Fc_+\LF \Hc \RF&  = \bigoplus_{i=0}^{\infty} \Hc_+^i = \mathbb{C}\oplus \Hc_{+1}\oplus \LF \Hc_{+1}\otimes \Hc_{+2}\RF\oplus \LF \Hc_{+1}\otimes\Hc_{+2}\otimes \Hc_{+3} \RF\oplus\cdots \\
		\Fc_{-}\LF \Hc \RF & = \bigoplus_{i=0}^{\infty} \Hc_{-}^i = \mathbb{C}\oplus \Hc_{-1}\oplus \LF \Hc_{-1}\otimes \Hc_{-2}\RF\oplus \LF \Hc_{-1}\otimes\Hc_{-2}\otimes \Hc_{-3} \RF\oplus\cdots\,, 
	\end{aligned}
\end{equation}
where $\Hc_{+n}$ and $\Hc_{-n}$ are n$^{\rm th}$ particle geometric superselection sectors. 
When we act the KG field operator on the total vacuum, we obtain 
\begin{equation}
	\hat{\phi} \vert 0\rangle_T  = \frac{1}{\sqrt{2}} \begin{pmatrix}
		\hat{\phi}_{-} & 0 \\ 
		0 & \hat{\phi}_{-} 
	\end{pmatrix}\begin{pmatrix}
		\vert 0\rangle_{+} \\ 
		\vert 0 \rangle_{-}
	\end{pmatrix} = \frac{1}{\sqrt{2}}\begin{pmatrix}
		\hat{\phi}_{+}\vert 0 \rangle_{+} \\ \hat{\phi}_{-}\vert 0 \rangle_{-}
	\end{pmatrix}
	\label{disumvac}
\end{equation}
This representation gives a creation of quantum field as a positive energy state at positions $\textbf{x}$  and $-\textbf{x}$ according to the positive energy state definition in vacuums $\vert 0\rangle_{+}$ and $\vert 0 \rangle_{-}$ respectively. Due to the fact that both $	\hat{\phi}_{+}\vert 0 \rangle_{+}$ and $	\hat{\phi}_{-}\vert 0 \rangle_{-}$ are supposed to describe a single quantum field (i.e., single degree of freedom) at two different spatial regions that are parity conjugates, we demand \eqref{newcomcond} to respect locality and causality. A generalization of \eqref{newcomcond} any operators corresponding to regions $+$ and $-$ should commute 
\begin{equation}
	[\Oc_{+}\LF t_p,\, \textbf{x} \RF,\, \Oc_{-}\LF -t_p^\prime, -\textbf{x}^\prime \RF] = 0\,. 
	\label{causalityeq}
\end{equation}
This is a new condition in addition to the usual demand of causality conditions that operators corresponding to space-like distances should vanish. Here in DQFT, it holds for both Fock space operators. 
\begin{equation}
	\begin{aligned}
		\left[\Oc_+\LF x \RF,\, \Oc_{+}(y)\right] & =0,\quad \LF x-y \RF^2>0\, \\
		\left[\Oc_{-}\LF -x \RF,\, \Oc_{-}(-y)\right] & =0,\quad \LF x-y \RF^2>0\,.
	\end{aligned}
	\label{comspace}
\end{equation} 
Recall that the Fock space of DQFT is a direct-sum of geometric superselection sectors $\Fc = \Fc_+\oplus \Fc_-$ leading to the description of quantum fields in parity conjugate regions of Minkowski space\footnote{Note that Parity operation takes a point at a radial distance $r$ to its antipode i.e., $\LF \theta,\,\varphi \RF\to \LF \pi-\theta,\,\pi+\varphi \RF$ which cannot be achieved by rotations.}. The two-point function of quantum field operators in DQFT is given by
\begin{equation}
\begin{aligned}
    \langle 0\vert \hat{\phi}\LF x \RF  \hat{\phi}\LF x^\prime \RF \vert 0\rangle & = \frac{1}{2} \langle 0_+\vert \hat{\phi}_+\LF x \RF  \hat{\phi}_+\LF x^\prime \RF \vert 0_+\rangle + \frac{1}{2} \langle 0_-\vert \hat{\phi}_-\LF -x \RF  \hat{\phi}_-\LF -x^\prime \RF \vert 0_- \rangle
    \end{aligned}
    \label{twopoint}
\end{equation}
Now, we can easily deduce the Feynman propagator, which is a time-ordered product of two field operators. According to this, the Feynman propagator of a quantum field between any two points in Minkowski becomes the sum of two terms
\begin{equation}
\begin{aligned}
    \langle 0\vert \tilde T \hat{\phi}\LF x \RF  \hat{\phi}\LF x^\prime \RF \vert 0\rangle & = \frac{1}{2} \langle 0_+\vert \tilde T \hat{\phi}_+\LF x \RF  \hat{\phi}_+\LF x^\prime \RF \vert 0_+\rangle + \frac{1}{2} \langle 0_-\vert  \tilde T\hat{\phi}_-\LF -x \RF  \hat{\phi}_-\LF -x^\prime \RF \vert 0_- \rangle
    \end{aligned}
    \label{twopointf}
\end{equation}
each describing the field component $\phi_\pm$ propagations in parity conjugate regions of physical space. The $\tilde T$ in \eqref{twopointf} denotes time ordering.

Note that $\hat{\phi}_{+}$ and $\hat{\phi}_{-}$ never get mixed by any interaction because of the definition of direct-sum \eqref{disum}. For example,
\begin{equation}
	\Lc_{int} = -\frac{\lambda}{3}\hat{\phi^3} = -\frac{\lambda}{3} \begin{pmatrix}
		\hat{\phi}_{+}^3 & 0 \\ 
		0 & 	\hat{\phi}_{-}^3 
	\end{pmatrix}
\end{equation}

As a natural consequence of $\Pc\Tc$  symmetry in Minkowski spacetime, all QFT computations extend seamlessly into the framework of DQFT without altering the results (for a detailed discussion, refer to \cite{Kumar:2023ctp,Kumar:2024oxf}). In the context of DQFT, the degrees of freedom in the standard model (SM), including particles $\vert SM\rangle$ and antiparticles $\vert \overline{SM}\rangle$ are represented through a direct-sum structure of the SM vacuum state:
\begin{equation}
    \vert 0_{SM}\rangle = \begin{pmatrix}
        \vert 0_{SM+}\rangle \\ 
        \vert 0_{SM-}\rangle 
    \end{pmatrix} \quad \vert SM\rangle = \frac{1}{\sqrt{2}}\begin{pmatrix}
        \vert SM_+\rangle \\ 
        \vert SM_-\rangle \end{pmatrix} \quad \vert \overline{SM}\rangle = \frac{1}{\sqrt{2}}\begin{pmatrix}
        \vert \overline{SM}_+\rangle \\ 
        \vert \overline{SM}_-\rangle 
    \end{pmatrix}
\end{equation}

The geometric superselection rule is applied consistently across all Fock spaces corresponding to the Standard Model's degrees of freedom, ensuring a uniform definition of parity-conjugated regions for every state. Although this work focused on the DQFT quantization of a real scalar field, the methodology extends seamlessly to complex scalars, fermions, and gauge fields. In DQFT, every quantum field is represented as a direct-sum of two components, which serve as $\Pc\Tc$ mirror images of one another, covering the entirety of Minkowski spacetime. As a result, the standard techniques of field quantization are straightforwardly adaptable to the DQFT framework.

\begin{itemize}
    \item Complex scalar field operator $\hat{\phi}_c  = \frac{1}{\sqrt{2}}\LF \hat{\phi}_{c\,+}\oplus  \hat{\phi}_{c\,-}\RF$ in DQFT is expanded as 
   \begin{equation}
   \begin{aligned}
    &  \hat{\phi}_{c\,\pm} = \int  \frac{d^3k}{\LF 2\pi \RF^{3/2}}\frac{1}{\sqrt{2\vert k_0\vert }}\Bigg[ a_{(\pm)\textbf{k}}e^{\pm ik\cdot  x}+b_{(\pm)\textbf{k}}^\dagger e^{\mp ik\cdot x}   \Bigg] \\ & \LT \hat{\phi}_{c\,+},\, \hat{\phi}_{c\,-} \RT =0\,,
    \end{aligned}
   \end{equation}
   where $a_{(\pm)\textbf{k}},\, a^\dagger_{(\pm)\textbf{k}}$ and $b_{(\pm)\textbf{k}},\, b^\dagger_{(\pm)\textbf{k}}$ are canonical creation and annihilation operators of the parity conjugate regions (denoted by subscripts $_{(\pm)}$) attached with geometric superselection sector.
   All the cross commutation relations of $a_{(\pm)},\, a^\dagger_{(\pm)}$ and $b_{(\pm)},\, b^\dagger_{(\pm)}$ vanish.  
   \item Fermionic field operator $ \hat \psi = \frac{1}{\sqrt{2}}\LF \hat \psi_+\oplus \hat \psi_- \RF$ in DQFT becomes 
   \begin{equation}
     \hat  \psi_{\pm} = \sum_{{\tilde s}} \int \frac{d^3k}{\LF 2\pi \RF^{3/2}\sqrt{2\vert k_0\vert}} \Bigg[ c_{{\tilde s}(\pm)\textbf{k}} u_{\tilde s}(\textbf{k}) e^{\pm ik\cdot x} + d_{{\tilde s}(\pm)\textbf{k}}^\dagger v_{\tilde s}(\textbf{k}) e^{\mp ik\cdot x}\Bigg]
   \end{equation}
where ${\tilde s}=1,2$ correspond to the two independent solutions of $\LF \slashed k+m\RF u_{\tilde s}=0$ and $\LF -\slashed k+m\RF v_{\tilde s}=0$ corresponding to spin-$\pm\frac{1}{2}$. The creation and annihilation operators of the Fock space geometric superselection sector, satisfy the anti-commutation relations $\Big\{ c_{{\tilde s}(\pm)\textbf{k}},\,c_{{\tilde s}(\pm)\textbf{k}}^\dagger \Big\}=1,\, \Big\{ c_{{\tilde s}(\mp)\textbf{k}},\,c_{{\tilde s}(\pm)\textbf{k}}^\dagger \Big\}=\Big\{ c_{{\tilde s}(\mp)\textbf{k}},\,c_{{\tilde s}(\pm)\textbf{k}} \Big\}=0$ leading to the new causality condition $\Big\{ \hat\psi_+,\,\hat \psi_-\Big\} =0$.
\item The vector field operator $\hat A_\mu = \frac{1}{\sqrt{2}}\LF \hat{A}_{+\mu}\oplus \hat A_{-\mu} \RF$ in DQFT expressed as 
\begin{equation}
	\hat{A}_{\pm \mu}= \int \frac{d^3k}{\LF 2\pi \RF^{3/2}\sqrt{2\vert k_0\vert }} e^{(\lambda)}_\mu\Bigg[ c_{(\pm \lambda)\textbf{k}} e^{\pm ik\cdot x}+c^\dagger_{(\pm \lambda)\textbf{k}} e^{\mp ik\cdot x}  \Bigg]
	\end{equation} 
	where $e^{(\lambda)}_\mu$ is the polarization vector satisfying the transverse and traceless conditions. The creation and annihilation operators $c_{(\pm \lambda)\textbf{k}},\,c_{(\pm \lambda)\textbf{k}}^\dagger$ satisfy the similar relations as \eqref{comcan}.
\end{itemize} 

All Standard Model calculations remain unchanged, as the interaction terms are decomposed into a direct-sum structure in the following manner. 
\begin{equation}
   \Lc_c \sim\Oc_{SM}^3=\begin{pmatrix}
        \Oc_{SM_+}^3 & 0 \\ 
        0 & \Oc_{SM_-}^3
    \end{pmatrix} \quad \Lc_q \sim \Oc_{SM}^4 = \begin{pmatrix}
        \Oc_{SM_+}^4 & 0 \\ 
        0 & \Oc_{SM_-}^4
    \end{pmatrix}
\end{equation}
Here, $\Oc_{SM}$ is an arbitrary operator involving any SM fields and their derivatives.\footnote{Remember that any derivative operators must be split into components joined by direct-sum operation.}  
The S-matrix also becomes the direct-sum with two parts
\begin{equation}
	\begin{aligned}
		S_T& = S_+ \oplus S_{-} 
	\end{aligned}
	\label{dQFTSM}
\end{equation}
where 
\begin{equation}
	S_+= T_1 \Bigg\{  e^{-i\int_{-\infty}^{\infty} H_{int} \,\, dt } \Bigg \},\quad S_{-}  = T_2 \Bigg\{ e^{i\int_\infty^{-\infty} H_{int } \,\,  dt } \Bigg \}
	\label{DQFTSM12}
\end{equation}
with $T_1,\, T_2$ representing the time orderings attached to the respective Fock space arrow of time.
Clearly, the DQFT framework leaves QFT calculations in Minkowski spacetime unaffected, as the spacetime itself is $\Pc\Tc$ symmetric. For instance, if we compute any scattering amplitude, such as the transition from N particles to M particles, the result remains the same under DQFT, namely
\begin{equation}
    A_{N\to M} = \frac{A^{N\to M}_+ \LF p_a, -p_b\RF  + A^{N\to M}_-\LF -p_a, p_b \RF}{2},\quad A^{N\to M}_+ \LF p_a, -p_b\RF = A^{N\to M}_-\LF -p_a, p_b \RF,
\end{equation}
where $p_a,\,p_b$ with $a=1,\cdots N$ and $b=1,\cdots M$ represent the 4-momenta of all the states involved in the scattering. 
$A_{\pm}$ represent amplitudes as a function of 4-momenta of initial and final states
computed in both vacuums $\vert 0_{SM\pm}\rangle$. Notice that the in (out) states in $\vert 0_{SM\pm}\rangle$ come with the opposite sign, which is due to the arrow of time being opposite in both the vacuums. The amplitudes $A_{\pm}$ are equal at any order in perturbation theory due to the $\Pc\Tc$ symmetry of Minkowski spacetime. 
The famous \cpt (charge conjugation, Parity, and Time reversal) invariance of scattering amplitudes \cite{Coleman:2018mew} also holds in both vacuums, which means
\begin{equation}
    A^{N\to M}_+(p_a, -p_b) = A^{M\to N}_+(-p_a, p_b) , \quad A^{N\to M}_-(-p_a, p_b) = A^{M\to N}_-(p_a, -p_b) \,.
\end{equation}
This is attributed to the fact that the \cpt operation of any scattering process would turn the outgoing anti-particles into in-going particles and vice-versa \cite{Coleman:2018mew}.  
  
In summary, we introduced a novel approach to quantum field theory by establishing a direct-sum mathematical framework that bridges $\Pc\Tc$-conjugate sheets of spacetime. Through the use of geometric superselection rules defined by parity-conjugate regions of physical space, we integrated the concept of two distinct arrows of time within a unified quantum state description. While DQFT preserves the practical outcomes of Standard Model particle physics, it offers a deeper insight into the role of "time" in quantum theory, providing a fresh perspective on its fundamental nature.

\subsection{Conformal diagram of DQFT in Minkowski spacetime}

\begin{figure}[h!]
	\centering
	\includegraphics[width=0.5\linewidth]{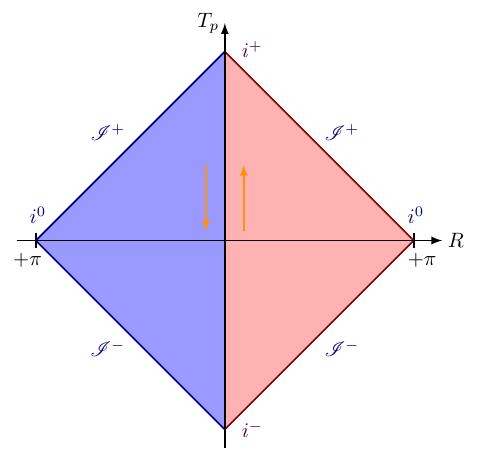}
	\caption{
		This figure is the depiction of	Minkowski spacetime in the DQFT construction, and we call it a "(quantum) conformal diagram" as it does not represent classical spacetime. The right (red) and left (blue) triangles are $\Pc\Tc$ conjugates of each other. The quantum field operator  $\hat{\phi}$ in DQFT is the direct-sum of $\hat{\phi}_+, \hat{\phi}_{-}$ defined on the right and left, respectively, which follows from \eqref{disum}.  The orange arrows represent the time direction.}
	\label{fig:minkowski-np}
\end{figure}

Minkowski spacetime \eqref{minst} is written in terms of the compactified coordinates by \cite{Griffiths:2009dfa}
\begin{equation}
	ds^2 = \frac{4}{\xi(T_p,R)}\LF-dT_p^2+ dR^2 + \sin^2(R) d\Omega^2 \RF\,, 
	\label{Minkowski-TR}
\end{equation}
where $-\pi+R <T_p<\pi-R$ and $0\leq R<\pi$, and $\xi(T_p,R) = \LF \cos T_p+\cos R\RF^2 $ is a positive function and $d\Omega^2$ represents the line element of two sphere. 
\begin{equation}
	\begin{aligned}
		T_p  & = \arctan\LF t_p+r \RF + \arctan\LF t_p-r \RF \\
		R & = 	\arctan\LF t_p+r \RF -\arctan\LF t_p-r \RF 
	\end{aligned}
	\label{TRplane}
\end{equation}
We can notice that under time reflection operation $\Tc$, we get
\begin{equation}
	\Tc: t_p\to -t_p \implies T_p\to -T_p, \quad R\to R\,.
\end{equation}
Fig.~\ref{fig:minkowski-np} represents the DQFT description of Minkowski spacetime, where we divide the space by Parity and assign opposite time evolution regions to the quantum fields in the spatial regions. 

We have several lessons from DQFT of Minkowski spacetime which we need to keep in mind for QFTCS 
\begin{itemize}
	\item There is no global arrow of time in Minkowski spacetime; in QFT, we follow a convention on the arrow of time $t_p: -\infty\to \infty$. 
	\item Direct-sum is a useful mathematical operation to represent the full Hilbert space of physical systems spanned by sub-physical regions with different arrows of time. On the other hand, direct-product (which is used to define entanglement in QM) is a mathematical operation of composite systems where the arrow of time is the same in both the sub-system and the environment. 
\end{itemize}

\section{Schwarzschild Black Hole and the discrete symmetries}

\label{sec:Schldmet}

The aim of this section is to discuss the dicrete spacetime transformations of SBH metric in different coordinates. Although this may be thought to be usual standard textbook material \cite{Griffiths:2009dfa}, our sincere appeal to the reader is that we specially pay attention to the discrete spacetime symmetries of SBH spacetime, which are pretty important for the understanding of quantum fields in such spacetime. Therefore, every small detail of this section is extremely important for the latter understanding and formulation of QFTCS. 

We can clearly notice that \eqref{SBHmet} is static, and it has the following discrete symmetries,
\begin{equation}
	\Tc: 	t\to -t,\quad \Pc: \LF \theta,\,\phi \RF \to \LF \pi-\theta,\,\pi+\varphi \RF 
	\label{SBHdis}
\end{equation}
which correspond to time reversal and parity operations. Notably, the $\Pc\Tc$ transformations \eqref{SBHdis} remain the same in the asymptotic Minkowski limit of the SBH metric \eqref{SBHmet} i.e., $r\to \infty$. 

But the form of metric \eqref{SBHmet} is not suitable everywhere to define quantum fields. It is because of the apparent coordinate singularity at $r=2GM$ which is certainly not a physical choice. Coordinate singularities are those where the curvature invariants are regular\footnote{For example, the Kretschman scalar $\Kc= R^{\mu\nu\rho\sigma}R_{\mu\nu\rho\sigma}= \frac{48G^2M^2}{r^6}$ for \eqref{SBHmet}.  }, thus, they are not the physical ones. An appropriate choice of coordinates is Kruskal-Szekers coordinates \eqref{KSmet} as we discussed in Sec.~\ref{sec:onetotwo}.

First of all, one may think that since \eqref{SBHmet} is spherically symmetric, common wisdom is to ignore angular coordinates $\LF \theta,\,\phi \RF$, which we strongly request our reader not to do so. 

\begin{figure}
	\centering
	\includegraphics[width=0.5\linewidth]{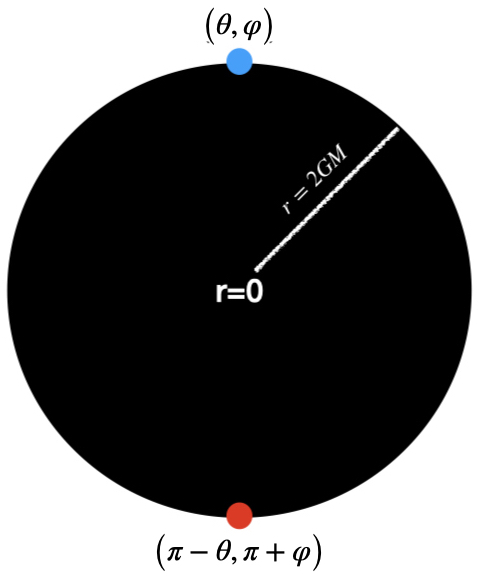}
	\caption{This picture depicts how the SBH \eqref{SBHmet} rip apart the parity conjugate points on the sphere of radius $r=2GM$ to space-like distances.}
	\label{fig:bhpics}
\end{figure}

Now, we aim to understand the nature of SBH spacetime \eqref{SBHmet}. One of its peculiarities is that the discrete set of points on the horizon
\begin{equation}
	\LF \theta,\,\phi \RF\Bigg\vert_{r=2GM}, \,\LF \pi-\theta,\, \pi+\varphi \RF \Bigg\vert_{r=2GM} 
\end{equation}
are space-like separated (see Fig.~\eqref{fig:bhpics}), and it takes an infinite amount of energy for anything to cross $r=0$ and reach the other point \cite{tHooft:2016rrl}. This explains why one must be careful when suppressing the angular coordinates in the SBH metric when one tries to make sense of spacetime quantum mechanically.
Following \eqref{rg2M} and \eqref{rg2M2}, the near-horizon approximation ($r_\ast \to -\infty$) gives
\begin{equation}
	r\to 2GM  \implies \Big\{u\to \infty,\, v\to -\infty\Big\} \implies \begin{cases}
		U\to 0^{\mp} & V\to 0^{\pm} \, \text{for}\quad r>2GM \\ 
		U\to 0^{\pm} & V\to 0^{\pm} \, \text{for}\quad r<2GM 
	\end{cases}
	\label{nearHlim}
\end{equation}

\subsection{An essential need to revise Hawking calculation}

The idea of this sub-section is to highlight the importance of revising Hawking's original calculation of 1975 \cite{Hawking:1975vcx}. Putting aside the remarkable result of Hawking's paper, the need to save unitarity is important, and, in fact, Hawking himself kept working on it for several decades \cite{Hawking:2014tga,Hawking:2015qqa}. Hawking pointed out, based on the success of AdS/CFT correspondence, that there should be a unitary evolution in (quantum) BH physics and he  investigated the subject in the context of BMS symmetries in collaboration with Perry and Strominger \cite{Hawking:2016msc,Strominger:2017zoo}. In the later Hawking works, the need for a careful analysis of the subject he initiated in 1975 with his seminal paper is suggested. 

\begin{figure}
	\centering
	\includegraphics[width=0.5\linewidth]{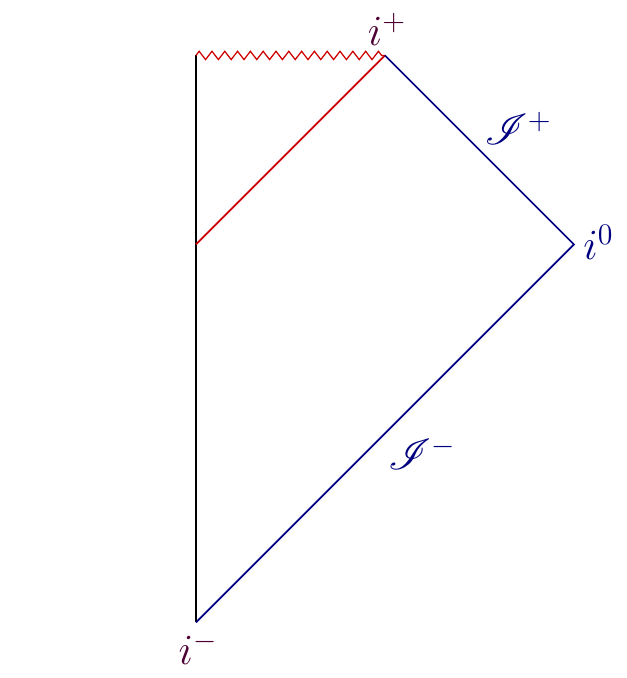}
	\caption{The figure depicts the (classical) Penrose diagram of BH formed by gravitational collapse. Given that the concept of time in quantum physics is entirely different from the classical concept (according to GR), making an assumption on "time" before quantization is the origin of the unitarity problem in BH physics.  }
	\label{fig:collapsingbh}
\end{figure}

As we discussed previously, in this section, there are discrete symmetries of SBH, and one must pay careful attention from a quantum mechanical point of view. We note that in the context of QFTCS, quantum gravity, there can be a concept of time with no classical analog as it is explicitly discussed in the book by Carlo Rovelli \cite{Rovelli:2004tv}. One indication for this would be that the famous Wheeler-De Witt (WDW) equation, the most basic equation in the scope of (quantum) GR, is grappled with the problem of time because the time variable does not explicitly appear in the equation \cite{Kiefer:2021zdq,Kiefer:2023bld,Chataignier:2023rkq}. 
Therefore, the unitarity problem arising from the Hawking calculation signals a caveat to us that we are missing notice of an unforeseen feature related to the concept of time in the basic assumptions that are made at the beginning of understanding (quantum) BHs. This is precisely what this paper is about. In the previous section, we noticed several definitions of KS coordinates \eqref{rg2M} and \eqref{rg2M2} that would leave the metric \eqref{KSmet} invariant. In the original Hawking calculations, only the first line of \eqref{rg2M} was taken into account and regarded as the BH formed by the gravitational collapse (the most popular classical Penrose diagram of it is depicted in Fig.~\eqref{fig:collapsingbh}). This is the most important primary caveat in Hawking's calculation because the classical and intuitive choice of coordinates and their values are put forward before quantization, regardless of the fact that "time" is a parameter in quantum theory. We know and understand so far the gravitational collapse from macroscopic, especially from GR point of view, and also some insights from microscopic physics in the sense at which point gravity would dominate from other fundamental forces of nature. But the whole process of collapse, formation of singularity (or formation of an object that closely resembles the BHs of GR \cite{Frolov:2014wja,Frolov:1998wf}) requires a clear understanding of scattering of fundamental degrees of freedom in the dynamical spacetime. It is because of this reason that several investigations have been studying the high energy scattering during the collapse, especially quantum gravity frameworks ask the question of ultra-high energy scattering of strings and formation of BHs \cite{tHooft:1987vrq,Giddings:2010pp,Nastase:2004pc,Chen:2021dsw,Gaddam:2020mwe,Gaddam:2020rxb}.  Thus, we perhaps require a full understanding of quantum gravity before we can certainly answer what microscopic physics is involved in the formation of BHs. But the question of unitarity, in the context of quantum fields in curved spacetime, takes precedence to the complete renormalizable theory of quantum gravity \cite{Giddings:2022jda}. 

Having said all these, we return to the (classical) assumption made by Hawking in his work \cite{Hawking:1975vcx}, which is restricting to a very particular definition of KS coordinates \eqref{KScoor} and cutting by hand all the discrete symmetries of the KS metric \eqref{KSmet} which are \eqref{rg2M} and \eqref{rg2M2}. This result of making a classical assumption of spacetime and fixing it as depicted in the (classical) Penrose diagram of BH formed by dynamical collapse (See Fig.~\ref{fig:collapsingbh}) is the origin of the unitarity problem. 
We have been institutionalized to think that time reversal in gravitational physics reverses whole dynamics, such as expanding the Universe to contracting the Universe and BH to the so-called White hole \cite{Griffiths:2009dfa}. 
However, one first needs to separate the concept of time classically from quantum mechanically. 
The classical understanding of spacetime only instructs us whether there is an event horizon or an apparent horizon (i.e., a horizon changing with time). 
In recent works, it has been shown how quantum fields can evolve backward in time in an expanding Universe \cite{Kumar:2023ctp,Gaztanaga:2024vtr,Kumar:2022zff,Gaztanaga:2024whs}. In the next section, we describe how quantum fields in SBH can respect its full discrete symmetries of KS metric \eqref{KSmet} and how unitarity emerges from there naturally. This follows from how we have constructed the DQFT of Minkowski spacetime, which is explicitly $\Pc\Tc$ symmetric from its construction of quantizing fields.

\section{Application of DQFT for Schwarzschild Black Hole}

\label{sec:DQFTBH}

This section uplifts the DQFT formalism to the context of SBH spacetime in KS metric, which is the suitable metric \eqref{KSmet} for quantization. The idea of DQFT based on the discrete spacetime transformations $\Pc\Tc$ makes use of the full description of the metric \eqref{KSmet} with all realizations in \eqref{rg2M} and \eqref{rg2M2}. The physical interpretation of this emerges accordingly, which we shall discuss in detail in the later parts of the section. Here, the BH with the horizon at $r=2GM$ can be regarded as a gravitational subsystem, and thus, we expand the KG field in terms of two sets of creation and annihilation operators corresponding to the interior and exterior of SBH (still restricting the approximation $r\approx 2GM$). To be more clear about this, let us start with a KG in the SBH spacetime 
\begin{equation}
	S_{KG}  = \frac{1}{2} \int \sqrt{-g}\, d^4x \phi \square \phi \,.
	\label{cact}
\end{equation}
Since the SBH spacetime \eqref{KSmet} is spherically symmetric, it is useful to express the KG scalar field in terms of Spherical harmonics 
\begin{equation}
	\phi \LF U,\, V,\,\theta,\phi \RF  = \frac{1}{\sqrt{2}} \sum_{\ell m} \frac{\phi_{\ell m}}{r} Y_{\ell m} \LF \theta,\,\varphi \RF
	\label{Sphscalar}
\end{equation}
Computing the d'Alembertian for the metric \eqref{KSmet} 
and integrating out on the $S_2$ sphere \eqref{cact} can be rewritten as 
\begin{equation}
	S_{KG} = \frac{1}{2} \int dUdV \phi_{\ell m}  \LF -\pd_U\pd_V + \frac{1}{r}\frac{\pd r}{\pd U\pd V}-\frac{A(r)\ell \LF \ell+1 \RF}{4r^2}  \RF \phi_{\ell m} 
	\label{KGeff}
\end{equation}
where we used the orthogonality of spherical harmonics
\begin{equation}
	\int_{0}^{2\pi} \int_{0}^{\pi} Y_{\ell m} Y_{\ell^\prime m^\prime} \sin\theta d\theta d\varphi = \delta_{\ell \ell^\prime} \delta_{mm^\prime}\,. 
\end{equation}
and $A(r) = \frac{2GM}{r}e^{1-\frac{r}{2GM}}$.
Note that our interest lies in the near horizon limits \eqref{nearHlim}, which means 
\begin{equation}
	\begin{aligned}
		\frac{1}{r}	\frac{\pd r }{\pd U\pd V} & = \frac{1}{UV} \frac{W\LT -\frac{UV}{4r_S^2} \RT }{\LF 1+ W\LT -\frac{UV}{4r_S^2}\RT\RF^4} \\ 
		& \approx -\frac{1}{4er_S^2}-\frac{5}{16e^2r_S^4}UV\cdots\,. 
	\end{aligned}
\end{equation}
which we can neglect together with the effective mass term in \eqref{KGeff} for the BHs whose mass is significantly heavier than the Planck mass $\sim 10^{-5}g$. 
Then, we are left with an effectively massless scalar field in 1+1 dimensions. Note that by using near horizon approximation and expanding the scalar field as \eqref{Sphscalar}, we obtained an effective scalar $\phi_{\ell m}$ which propagates effectively in two-dimensional spacetime. We may be tempted to see it as local Minkowski spacetime, but we must distinguish it from the asymptotic Minkowski spacetime of \eqref{SBHmet} in the limit $r\to \infty$. 

To quantize the scalar $\phi_{\ell m} (U,\,V)$ which is a field component of the original field $\phi$ (See \eqref{Sphscalar}) we need to promote it as an operator. Since we are quantizing a relativistic spherical wave  \eqref{Sphscalar}, in the SBH spacetime \eqref{KSmet}, we must not think about it in terms of (localized) particles, a common practice in the context of QFT of Minkowski spacetime.  Nevertheless, the concept of a particle in curved spacetime is a subject of separate investigation \cite{Colosi:2004vw}.

It is well-known from the classical understanding of SBH that the concept of time is totally different inside and outside the Schwarzschild radius  $r_S= 2GM$, in the sense that light cones of a classical point-like object would get rotated by 90\textdegree at $r_S= 2GM$. Also, it has been shown that the exterior Schwarzchild metric is static, whereas the interior metric is related to Kantowski-Sachs, which is a cosmological metric \cite{Doran:2006dq}. 
Before we do any quantization, we must take every observation and lesson from classical physics. Then, when we do quantization, we must respect aspects of discrete symmetries and leave aside the intuitions borrowed from classical physics. Since the nature of the metric in the BH interior and exterior is different (the concept of time is different), we apply the geometric superselection rule when we quantize the field.
Now, what we do is to separate the spatial region of SBH \eqref{KSmet} by the horizon 
\begin{equation}
	\hat{\Phi}  = \hat{\Phi}_{int}  \oplus \hat{\Phi}_{ext} \,, 
	\label{splitei}
\end{equation}
where
\begin{equation}
\Phi = \hat{\phi}_{\ell m},\quad 	\hat{\Phi}_{int} = \hat{\phi}_{\ell m} \Bigg\vert_{r\lesssim 2GM},\quad    \hat{\Phi}_{ext} = \hat{\phi}_{\ell m} \Bigg\vert_{r\gtrsim 2GM}\,, 
	\label{fopio}
\end{equation}
are the quantum fields defined for the interior and exterior {(with the notation $r\lesssim 2GM$ and $r \gtrsim 2GM$ indicating near horizon approximation)} of the SBH \eqref{SBHmet} represented by the KS metric \eqref{KSmet}. 
The KS metric has (discrete) symmetries that are intrinsic to quantum theory (once more, we recall that time is a parameter in quantum theory). The exterior of the SBH  \eqref{SBHmet} is given by \eqref{rg2M}, while the interior of the SBH \eqref{SBHmet} is described by \eqref{rg2M2}. Both in the interior and exterior, we have the discrete symmetries \eqref{dissym},  which we use to formulate a direct-sum QFT for the interior and exterior regions of SBH. Therefore, we associate the above field operators to the two Fock spaces that describe the interior and exteriors of SBH. The direct-sum of these Fock spaces shall "bridge" the interior and the exterior. 
\begin{equation}
	\Fc_{BH} = \Fc_{int}\oplus \Fc_{ext} \,.  
	\label{fockBH}
\end{equation}
This is exactly the place where we differ from Hawking's construction which uses a normal summation, i.e., $\hat{\phi}_{\ell m} = \hat{\phi}_{\ell m}^{ext}+\hat{	\phi}_{\ell m}^{int}$ See. Eq. (2.7) of Hawking paper \cite{Hawking:1976ra}. The simple change of normal summation to direct-sum operation has profound implications in resolving the problem of unitarity, which we shall discuss in  Sec.~\ref{sec:NotPen}. 

Quantum theory is built on defining commutation and non-commutation relations, for example, \eqref{newcomcond}, \eqref{canDQFT}, \eqref{causalityeq} and \eqref{comspace} in the context of Minkowski DQFT. The most important lesson from QFT construction in Minkowski spacetime is that the causality condition, which means the operators corresponding to spacelike distances, must vanish. In the context of SBH, we need to impose the operators corresponding to the space-like separated points (which are Parity conjugates. See Fig.~\ref{fig:bhpics}) must commute. We shall show how one can achieve this in a while. 

Then another question that arises here is what should be the commutator of
\begin{equation}
	\Big[\hat{	\Phi}_{int},\, \hat{\Phi}_{ext}\Big] =?\,. 
	\label{qcom}
\end{equation}
According to Hawking's conjecture, the above commutator is zero (See Eq.~(2.8) of \cite{Hawking:1976ra}). {Note that our terminology of interior ($r\lesssim 2GM$) and exterior ($r\gtrsim 2GM$) parts, of the quantum field, corresponds to "ingoing" and "outgoing" modes in the Hawking's paper (in the near horizon approximation) \cite{Hawking:1976ra}.}
So Hawking assumes that the quantum fields created in the SBH interior (just inside the horizon) are totally independent of what is created in the exterior (just outside the horizon), and it is exactly the origin of the information problem. This has truly emerged from the "classical" intuition of BH no hair theorem according to classical GR \cite{Misner:1973prb,Sotiriou:2015pka}. If the information paradox is to be solved or at least the first solution towards it has to emerge from the non-zero commutation of \eqref{qcom} which can be determined by a rather remarkable calculation of Dray and 't Hooft on the gravitational backreaction of (classical) point particles in GR \cite{Dray:1984ha}. We discuss this point further, based on the recent calculations of 't Hooft \cite{tHooft:1996rdg,tHooft:2016rrl,tHooft:2016qoo,tHooft:2021lyt,tHooft:2015pce,tHooft:2018fxg,tHooft:2022bgo,tHooft:2022umh,tHooft:2023ggx},  which will help to determine \eqref{qcom}.

\subsection{Gravitational backreaction and quantum algebra}

\label{sec:shapiro}

QFT has emerged with a clear understanding of classical special relativity and quantum mechanics. Both are brought together by a simple demand of commuting operators for space-like distances \eqref{comspace}. Similarly, when we speak about quantum states created at $r\lesssim 2GM$ and $r\gtrsim 2GM$ in the context of BH evaporation\footnote{It is well-known that Hawking quanta is the result of particle pair creation at $r\lesssim 2GM$ and $r\gtrsim 2GM$ and it is the effect of so-called stretched horizon as explained in detail by Mathur \cite{Mathur:2009hf}. It is worth mentioning the seminar work of Starobinsky \cite{Starobinskii:1973hgd}, a precursor to Hawking's paper, already mentions boson pair creation at the BH horizon.  } \cite{Mathur:2009hf,Mathur:2011uj,Calmet:2022swf}, it is essential to know from GR that if a (classical) particle at $r\lesssim 2GM$ can gravitationally affect its counterpart at  $r\gtrsim 2GM$. The answer is 'Yes' and is called the gravitational backreaction effect due to the shock waves created by (classical) particles in motion at the exterior and interior of the SBH (near horizon). This is the most famous calculation by Dray and 't Hooft \cite{Dray:1984ha} (See also works by Lousto and Sanchez \cite{Lousto:1988sp}), which is well-summarized in Sec.~2 of \cite{Betzios:2016yaq}. The calculation of Dray and 't Hooft has been the central subject of investigation over the years in the context of gravitational memory effects, ultra-Planckian energy scatterings, gravitational collapse, and ultimately the formation of BHs \cite{Gray:2021dfk,Giddings:2001bu,Buonanno:2022pgc,Lam:2018pvp,He:2023qha,deVega:1988wp,Amati:1992zb,Kabat:1992tb,Eardley:2002re,Donnay:2018ckb,Strominger:2017zoo,Lousto:1991xb}.
Despite the importance of Dray-'t Hooft' s paper highlighted in these later investigations about BHs, Dray-'t Hooft´s result actually modifies the way we understand (quantum) BHs\footnote{In this context 't Hooft in recent years emphasized the role of gravitational backreaction in understanding Hawking radiation \cite{tHooft:2016qoo,tHooft:2022bgo} and presented his new interpretation of (quantum) BH with antipodal identification \cite{SANCHEZ1987605} (i.e., classical topological change \cite{tHooft:2016rrl}) and later the reformed proposal with quantum clones \cite{tHooft:2022bgo}. We do not endorse these interpretations because they involve quite unusual boundary conditions and create problems for understanding the collapse and evaporation of BHs. Furthermore, the 't Hooft proposal completely removes the discussion about SBH interior and claims it is unimportant. However, we argue that the interior region of SBH is very important, and it plays a non-trivial role in the way we understand unitary BH radiation. The importance of BH interior from a quantum mechanical point of view is shown in recent works by Norma G. Sanchez \cite{Sanchez:2018lnb,Sanchez:2020rqj,Sanchez:2023ylf}
	  }. The remarkable efforts of 't Hooft on the subject of gravitational backreaction and the corresponding quantum algebra \cite{tHooft:1996rdg,tHooft:2016rrl,tHooft:2016qoo,tHooft:2021lyt,tHooft:2015pce,tHooft:2016sdu,tHooft:2022bgo,tHooft:2023ggx,tHooft:2022umh} together with our DQFT formalism give us the more complete and concrete picture of (quantum) BHs. Below we shall discuss briefly the result of Dray-'t Hooft paper \cite{Dray:1984ha} and also a few relevant parts of recent 't Hooft' s remarkable non-commutative algebra associated with positions and momenta of the quantum mechanical particles in the interior and exterior of SBH. In this paper, we explain this within our context and notations to avoid confusion. 

Note that we are interested in the near horizon limits given by \eqref{nearHlim} and how a (classical) massless particle at $r\lesssim 2GM$ can gravitationally affect the (classical) massless particle at $r\gtrsim 2GM$. This is the meaning of gravitational backreaction. Understanding this would tell us about the RHS of \eqref{qcom}. Below, we apply the results of gravitational backreaction and the corresponding quantum algebra to our context \cite{Dray:1984ha,Betzios:2016yaq,Nastase:2004pc,tHooft:2016qoo}. 

Suppose we have a massless (classical) point particle with momentum (density) $P_i$ at an angle $\Omega$ moving in the direction of V 
near the horizon $r\lesssim 2GM$, it creates a shockwave $f\LF \Omega,\, \Omega^\prime \RF$ in the transverse direction of its motion  described by  \cite{Dray:1984ha,Betzios:2016yaq,Nastase:2004pc}
\begin{equation}
	ds^2 \approx - 16G^2M^2 \LF  d\bar Ud\bar V + \delta(\bar U)   f\LF\Omega,\,\Omega^\prime\RF d\bar U^2\RF  + \LF 2GM \RF^2 d\Omega^2,
	\label{shokm}
\end{equation}
where $\bar U = \frac{U}{4GM},\, \bar V=\frac{V}{4GM}$ are dimensionless KS coordinates.
The delta function in the metric $\delta(\bar U)$ signifies the location of the source at $\bar U\approx 0$ corresponding to the energy-momentum tensor 
\begin{equation}
	T_{\bar U\bar U} = P_i \delta(\bar U) \delta(\Omega) 
\end{equation}
Here $P_i$ is(ingoing) momentum density which we rescale it with $\bar{P}_i = \frac{P_i}{r_S^4}$. 

The function $f(\Omega,\, \Omega^\prime)$ is determined by solving the GR equations of motion in the near horizon approximation, and it gives
\begin{equation}
	(	\Delta_S-1 )f(\Omega,\,\Omega^\prime) = -16\pi G r_S^{-2} e^{-1} \bar P_i\delta^2\LF  \Omega,\,\Omega^\prime  \RF
\end{equation}
where $\Delta_S$ is the sphere Laplacian, $\delta^2\LF  \Omega,\,\Omega^\prime  \RF$, $\Omega\equiv \LF \theta,\,\varphi \RF,\, \Omega^\prime\equiv \LF \theta^\prime,\,\varphi^\prime \RF$ is the delta function. It was found that the massless (classical) point particle at the position $\bar V_e$ (at an angle $\Omega^\prime$) near the horizon ($r\lesssim 2GM$) 
experience the shockwave \eqref{shokm} (backreaction), and consequently,  its position gets shifted by an amount given by
\begin{equation}
	\bar V_e\to \bar V_e+ f\LF \Omega,\, \Omega^\prime \RF
	\label{shiftout}
\end{equation}
Now if we perform a similar exercise with a massless particle with momentum $P_e$  moving along the direction U near the horizon ($r\gtrsim 2GM$) at an angle $\Omega$, then it creates its own gravitational field with a shock wave, it causes a shift in the position of the massless (classical) point particle at the position $\bar U_i$ near the horizon ($r\lesssim 2GM$)
\begin{equation}
	\bar U_i\to \bar U_i+ f\LF \Omega,\, \Omega^\prime \RF\,. 
	\label{shiftin}
\end{equation}
Since we are speaking about the near horizon approximation ($\vert U\vert\to 0,\,\vert V\vert \to 0$) \eqref{nearHlim}, we approximate $\bar V_e = \bar U_i\approx 0 $ before the gravitational shift and rename the new positions as $\bar V_e$ and $\bar U_i$.  Following \cite{tHooft:2016qoo} we express
\begin{equation}
    \bar V_e = \frac{8\pi G}{r_S^2}\int d^2\Omega^\prime \tilde{f}\LF \Omega,\,\Omega^\prime \RF \bar{P}_i,\quad \bar U_i = -\frac{8\pi G}{r_S^2}\int d^2\Omega^\prime \tilde{f}\LF \Omega,\,\Omega^\prime \RF \bar{P}_e
\end{equation}
where $\tilde{f}\LF \Omega,\,\Omega^\prime \RF$ satisfies $\LF \Delta_S -1\RF \tilde{f} = -\delta^{2}\LF \Omega,\,\Omega^\prime \RF$.
We can expand positions $\LF \bar V_e,\,\bar U_i\RF$ and momenta $\LF \bar P_e,\,\bar P_i\RF$ with the spherical harmonics as
\begin{equation}
	\begin{aligned}
\bar V_e & \to \sum_{\ell,\, m} \bar V_e Y_{\ell m}\LF \theta,\,\phi \RF,\quad \bar U_i \to  \sum_{\ell,\, m} \bar U_i Y_{\ell m}\LF \theta,\,\phi \RF \\
		\bar P_e & \to \sum_{\ell,\, m} \bar P_e Y_{\ell,\, m}\LF \theta,\,\phi \RF,\quad \bar P_i \to  \sum_{\ell m} \bar P_i Y_{\ell,\, m}\LF \theta,\,\phi \RF
	\end{aligned}
\end{equation}
As a result of the above steps, we can map the position of the massless point particle $\bar V_e$ with the momentum of the massless (classical) particle $\bar P_i$ and similarly map $\bar U_i$ with $\bar P_e$ as 
\begin{equation}
	\bar V_e = \frac{8\pi G}{r_S^2 \LF \ell^2+\ell +1 \RF}\bar P_i,\quad \bar U_i = -\frac{8\pi G}{ r_S^2\LF \ell^2+\ell +1 \RF}\bar P_e\,. 
	\label{inoeff}
\end{equation}

\begin{figure}
	\centering
	\includegraphics[width=0.7\linewidth]{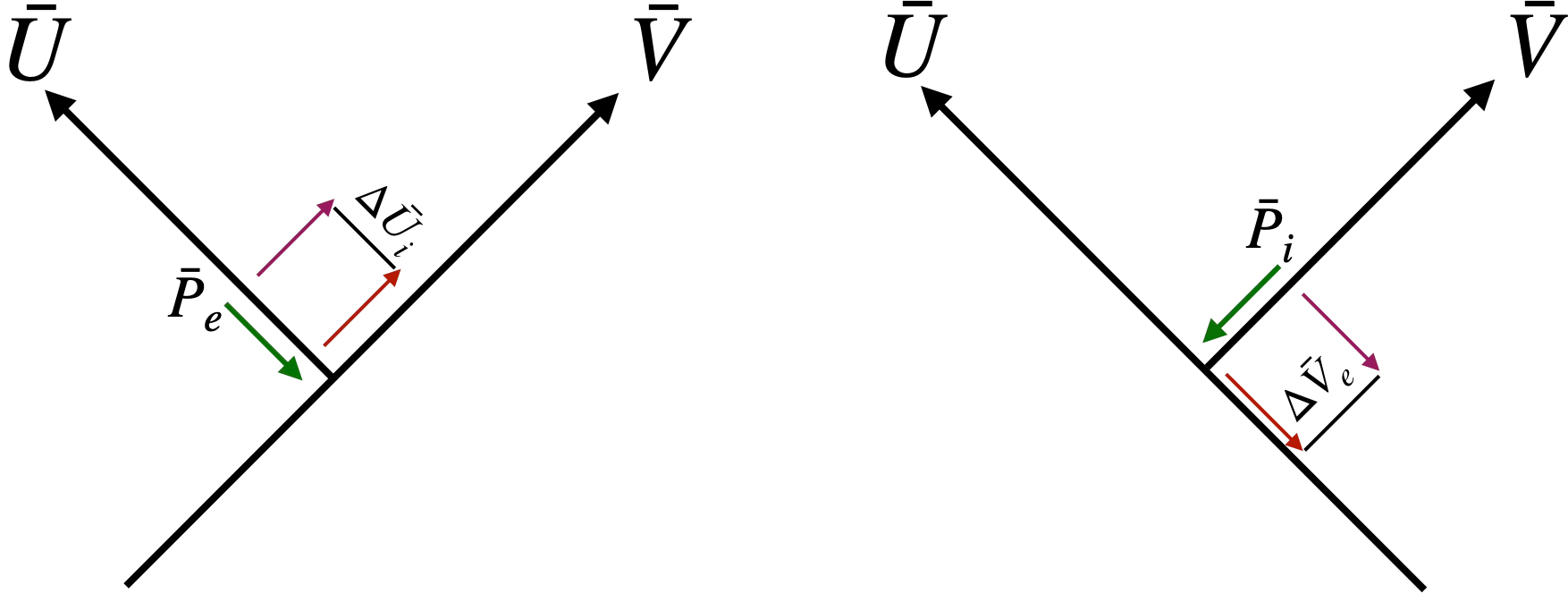}
	\caption{In the above picture, we depict the SBH interior and exterior particles affecting each other gravitationally. An interior ($r\lesssim 2GM$) (classical) particle with momentum $\bar P_i$ gravitationally cause a shift $\Delta \bar V_e = f(\Omega,\,\Omega^\prime)$ in the position of the exterior outgoing ($r\gtrsim 2GM$) particle at $\bar V_e$. Similarly, an exterior (classical) particle with momentum $\bar P_e$ causes gravitationally a shift $\Delta U_i=f(\Omega,\,\Omega^\prime)$ in the of the interior classical particle position $\bar U_i$. Applying position and momentum uncertainty relations \eqref{uncert} we arrive at the quantum backreaction algebra \eqref{noncompm} that exposes the quantum gravitational connection between interior and exterior quantum states.     }
	\label{fig:gbr}
\end{figure}

Note that in the above steps, we ignore all the effects of transverse momenta since for sufficiently large spherically symmetric SBH, the transverse effects can be negligible \cite{tHooft:2022bgo,tHooft:2023ggx}. The gravitational backreaction of the interior state on the exterior state, and vice versa, can be schematically depicted as in Fig.~\ref{fig:gbr}.

Given now that GR clearly teaches us how a (classical) massless particle moving along the direction of V can gravitational affect the particle moving along the direction of U through gravitational backreaction due to shock waves (it is also known as the Shapiro effect), 't Hooft has applied the rules of quantum mechanics to find the following commutation relations which form the basis of BH QFT construction and ultimately allow us to find an answer to our question \eqref{qcom}. 
\begin{equation}
	\begin{aligned}
	\Big[\hat{\bar V}_e,\, \hat{\bar P}_e\Big] = i\hbar,\quad 	\Big[\hat{\bar U}_i,\, \hat{\bar P}_i\Big] = i\hbar
	\end{aligned}
	\label{uncert}
\end{equation}
As a consequence of \eqref{inoeff} we end up with a most curious relation that forms the basis for the combined effect of gravity and quantum mechanics 
\begin{equation}
	\begin{aligned}
		\Big[\hat{\bar U}_i,\, \hat{\bar V}_e\Big] = i\hbar \frac{8\pi G}{r_S^2\LF \ell^2+\ell+1 \RF},\quad 	\Big[\hat{\bar P}_i,\, \hat{\bar P}_e\Big] = i\hbar  \frac{r_S^2\LF \ell^2+\ell+1 \RF}{8\pi G}
	\end{aligned}
	\label{noncompm}
\end{equation}

Recalling that QFT is nevertheless an extension of quantum mechanics with relativistic concepts \cite{Coleman:2018mew}, we can already read the RHS of \eqref{qcom} from the RHS of \eqref{noncompm}. The rather very curious thing about the relations \eqref{noncompm} is that it involves a combination of $G,\, \hbar,\, c$ and also Schwarzschild radius $r_S$, which signals that BHs are the key to understanding quantum gravity \cite{tHooft:2016sdu}.

\subsection{BH DQFT, unitary Hawking radiation, and conformal diagram}
\label{sec:NotPen}

In \eqref{splitei} and \eqref{fockBH}, we have the field operator and the respective Fock space written as a direct-sum of the two components corresponding to the exterior and interior of the SBH \eqref{SBHmet} i.e.,\, $ r\gtrsim 2GM$ and $r\lesssim 2GM$. Now, in a similar way to how we have DQFT of Minkowski spacetime, based on the discrete symmetries of the Minkowski metric (Sec.~\ref{sec:revMin}), we can formulate BH DQFT based on the discrete spacetime transformations \eqref{dissym} both for the interior and exterior of SBH as follows. 

In the view of  \eqref{dissym}, \eqref{rg2M} and \eqref{rg2M2} and BH DQFT, we expand \eqref{splitei} as 
\begin{equation}
	\begin{aligned}
		\hat{\Phi}  & = \hat{\Phi}_{int} \oplus \hat{	\Phi}_{ext} \\ 
		& = \frac{1}{\sqrt{2}} \LF  \hat{\Phi}^{I}_{int} \oplus \hat{	\Phi}^{II}_{int} \RF \oplus \frac{1}{\sqrt{2}}\LF  \hat{\Phi}^{I}_{ext} \oplus \hat{	\Phi}^{II}_{ext} \RF\, 
	\end{aligned}
\end{equation}
where 
\begin{equation}
	\begin{aligned}
		\hat{\Phi}^{I}_{ext} & = \int_{-\infty}^{\infty} \frac{dk}{\sqrt{2\pi}} \frac{1}{\sqrt{2\vert k\vert }} \Bigg[ \hat{b}_{I k} e^{-i\vert k\vert T+ikX} + \hat{b}_{I k}^\dagger e^{i\vert k\vert T-ikX}  \Bigg] \\
		& = \int_{0}^{\infty} \frac{d\omega}{\sqrt{2\pi}} \frac{1}{\sqrt{2\omega }} \Bigg[ \hat{b}_{I \omega} e^{-i\omega U} +  \hat{b}_{I -\omega} e^{-i\omega V} +  \hat{b}_{I \omega}^\dagger e^{i\omega U} + \hat{b}_{I -\omega}^\dagger e^{i\omega V}  \Bigg]  \\
		\hat{\Phi}^{II}_{ext} & = \int \frac{dk}{\sqrt{2\pi}} \frac{1}{\sqrt{2\vert k\vert }} \Bigg[ \hat{b}_{II k} e^{i \vert k\vert T-ikX} + \hat{b}_{II k}^\dagger e^{-i\vert k\vert T+ikX}  \Bigg] \\
		& =  \int_{0}^{\infty} \frac{dk}{\sqrt{2\pi}} \frac{1}{\sqrt{2\omega }} \Bigg[ \hat{b}_{II \omega} e^{i\omega U} +  \hat{b}_{II -\omega} e^{i\omega V} +  \hat{b}_{II \omega}^\dagger e^{-i\omega U} + \hat{b}_{II -\omega}^\dagger e^{-i\omega V}  \Bigg] \\
		\hat{\Phi}^{I}_{int} & = \int \frac{dk}{\sqrt{2\pi}} \frac{1}{\sqrt{2\vert k\vert }} \Bigg[ \hat{c}_{I k} e^{i \vert k\vert T-i kX} + \hat{c}_{I k}^\dagger e^{-i \vert k\vert T+ikX}  \Bigg] \\
		\hat{\Phi}^{II}_{int} & = \int \frac{dk}{\sqrt{2\pi}} \frac{1}{\sqrt{2\vert k\vert }} \Bigg[ \hat{c}_{II k} e^{-i \vert k\vert T+ikX} + \hat{c}_{II k}^\dagger e^{i\vert k\vert T-ikX}\Bigg]
	\end{aligned}
	\label{BHdecom}
\end{equation}
where $\omega= \vert k \vert $. 
Here 
\begin{equation}
	\begin{aligned}
	T  = \frac{U+V}{2\sqrt{e}},\quad 	X = \frac{V-U}{2\sqrt{e}}, \quad r>2GM \\ 
	X = \frac{U+V}{2\sqrt{e}},\quad T = \frac{V-U}{2\sqrt{e}},\quad r<2GM\,.
	\end{aligned}
	\label{TX}
\end{equation}
The creation and annihilation operators above satisfy the following commutation relations
\begin{equation}
	\begin{aligned}
		\Big[\hat{b}_{I\,k},\,\hat{b}_{I\,k^\prime}^\dagger\Big] & = 		\Big[\hat{b}_{II\,k},\,\hat{b}_{II\,k^\prime}^\dagger\Big] = \hbar \delta^{(3)}\LF k-k^\prime \RF\\
		\Big[\hat{b}_{I\,k},\,\hat{b}_{II\,k^\prime}\Big] &=	\Big[\hat{b}_{I\,k},\,\hat{b}_{II\,k^\prime}^\dagger\Big] = \Big[\hat{b}_{I\,k}^\dagger,\,\hat{b}_{II\,k^\prime}^\dagger\Big]  =0\,.\\
		\Big[\hat{c}_{I\,k},\,\hat{c}_{I\,k^\prime}^\dagger\Big] & = 	\Big[\hat{c}_{II\,k},\,\hat{c}_{II\,k^\prime}^\dagger\Big] = \hbar \:\: \delta^{(3)}\LF k-k^\prime \RF\\
		\Big[\hat{c}_{I\,k},\,\hat{c}_{II\,k^\prime}\Big] &=	\Big[\hat{c}_{I\,k},\,\hat{c}_{II\,k^\prime}^\dagger\Big] = \Big[\hat{c}_{I\,k}^\dagger,\,\hat{c}_{II\,k^\prime}^\dagger\Big]  =0\,.
	\end{aligned}
	\label{Bhcomre}
\end{equation}
Here the 2nd and 4th line of commutation relations \eqref{Bhcomre} leads to
\begin{equation}
	\begin{aligned}
	\Big[ \hat{\Phi}_{I ext},\, \hat{\Phi}_{II ext}  \Big]  = 0 \\ 
		\Big[ \hat{\Phi}_{I int},\, \hat{\Phi}_{II int}  \Big]  = 0\,. 
		\end{aligned}
		\label{extcom}
\end{equation}
which is similar to the one \eqref{causalityeq}. This condition takes care of operators at space-like distances who should commute (See Fig.~\ref{fig:bhpics}).

As a consequence of \eqref{noncompm}, the exterior and interior quantum field operators should be non-commutative as
\begin{equation}
	\begin{aligned}
	\Big[ \hat{\Phi}_{I int},\, \hat{\Phi}_{I ext}  \Big] &   = i\hbar \frac{8\pi G}{r_S^2\LF \ell^2+\ell+1 \RF} \\
		\Big[ \hat{\Phi}_{II int},\, \hat{\Phi}_{II ext}  \Big] &   = i\hbar \frac{8\pi G}{r_S^2\LF \ell^2+\ell+1 \RF} \,. 
	\end{aligned}
 \label{intextf}
\end{equation}
 {It is worth to notice that in the limit $r_S\to \infty$ the commutators \eqref{intextf} tend to vanish. The same thing happens in the limit $\ell\gg 1$. This would mean the bigger the black hole, the gravitational backreaction effects tend to be negligible. The low-$\ell$ modes are key for understanding quantum effects BH horizon as it was also highlighted by recent related investigations \cite{Gaddam:2020mwe}. }

\begin{figure}
	\centering	\includegraphics[width=0.7\linewidth]{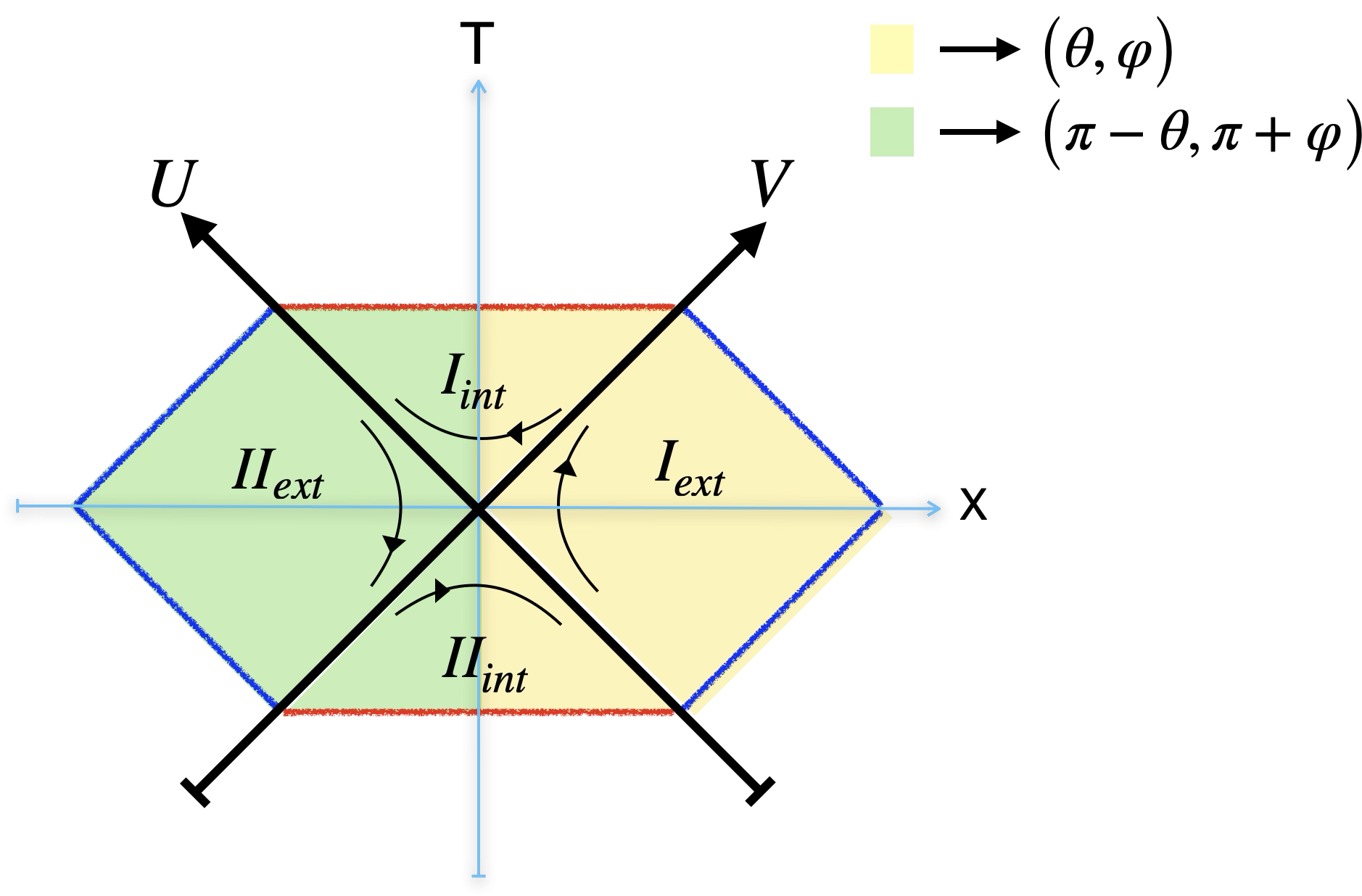}
	\caption{This represents the conformal diagram of BH DQFT. Here $I_{ext}$ describes an exterior region of SBH spanned $\LF \theta,\,\varphi \RF$ where a component of the quantum field $ \hat{	\Phi}_{I\, ext}\vert 0\rangle_{I ext}$ evolves forward in time while the  $II_{ext}$ described the exterior region spanned by the parity conjugate points  $\LF \pi-\theta,\,\pi+\varphi \RF$ where the second component $\hat{	\Phi}_{II\, ext}\vert 0\rangle_{II ext}$  of quantum field evolves backward in time. Similarly, $I_{int}$ describes interior region of SBH where $ \hat{	\Phi}_{I\, int}\vert 0\rangle_{I ext}$ evolves forward in time while the  $II_{int}$ describes the interior region where the second component of quantum field $\hat{	\Phi}_{II\, int}\vert 0\rangle_{II int}$ evolves backward in time. The arrows represent the direction of the time evolution of a quantum field in each region.  Every point in the yellow shared area represents the points $\LF \theta,\,\varphi \RF$, whereas the green shared regions represent the parity conjugated points $\LF \pi-\theta,\,\pi+\varphi \RF$.
 The red line represents $r=0$ singularity and is identified in the upper and lower half of the interior regions.   }
	\label{fig:fig4v3}
\end{figure}

So \eqref{intextf} and \eqref{extcom2} imply
\begin{equation}
	\begin{aligned}
		\Big[\hat{c}_{I\,k},\,\hat{b}_{I\,k^\prime}\Big] & = 	\Big[\hat{c}_{II\,k},\,\hat{b}_{II\,k^\prime}\Big]= 	i\hbar \frac{8\pi G}{r_S^2\LF \ell^2+\ell+1 \RF} \delta^{(3)}\LF k-k^\prime \RF,\quad \Big[c_{I\,k},\,b^\dagger_{I,\,k^\prime}\Big] = 0  \\
		\Big[\hat{c}_{I\,k},\,\hat{b}_{II\,k^\prime}\Big] &=	\Big[\hat{c}_{I\,k},\,\hat{b}_{II\,k^\prime}^\dagger\Big] = \Big[\hat{c}_{I\,k}^\dagger,\,\hat{b}_{II\,k^\prime}^\dagger\Big]  =0\,. 
	\end{aligned}
	\label{infcomm}
\end{equation}
In the near-horizon limit \eqref{nearHlim} the quantum vacuum of SBH as per the decomposition \eqref{BHdecom} is a direct-sum as (See Fig.~\ref{fig:fig4v3})
\begin{equation}
	\begin{aligned}
	\vert 0\rangle_{BH} & = \Big( \vert 0 \rangle_{I int} \oplus \vert 0 \rangle_{II int} \Big) \oplus  \Big( \vert 0 \rangle_{I ext} \oplus \vert 0  \rangle_{II ext} \Big) 
	\end{aligned}
	\label{BHvac}
\end{equation}
where the components of vacuum \eqref{BHvac} defined by
\begin{equation}
	\begin{aligned}
		\hat{b}_{I k} \vert 0 \rangle_{I ext} &  = 0,\quad	\hat{b}_{II k} \vert 0 \rangle_{II ext}  = 0 \\ 
		\hat{c}_{I k} \vert 0 \rangle_{I int} &  = 0,\quad	\hat{c}_{II k} \vert 0 \rangle_{II int}  = 0\,. 
	\end{aligned}
 \label{vacBHop}
\end{equation}
{Note that the decomposition of field components in \eqref{Bhcomre} is done with an approximation that spacetime is locally "flat" \eqref{nearflat}. This means the vacuum is chosen by imposing a Hadamard condition \cite{Birrell:1982ix} such that an infalling observer would not see any quantum states or firewalls at the horizon. Since our vacuum structure is a direct-sum of two, each describing parity conjugate regions with opposite arrows of time, we impose Hadamard conditions to match the direct-sum Minkowski vacuum \eqref{tFs}. }

The total Fock space is again a direct-sum corresponding to the different spatial regions and arrows of time ($r\lesssim 2GM$ and $r\gtrsim 2GM$)
\begin{equation}
	\Fc_{BH} = \Fc_{I int}\oplus \Fc_{II int}\oplus  \Fc_{I ext} \oplus \Fc_{II ext}
\end{equation}
Thus, there are 4 geometric superselection sectors because there are 4 discrete transformations \eqref{rg2M} and \eqref{rg2M2}. 

In all the above expressions, the following meaning of notations is implicit 
\begin{itemize}
	\item Region $I_{int}$ is the interior spatial region of SBH $r\lesssim 2GM$ with $U>0,\, V>0$ i.e., half of the interior. The vacuum in this region is $\vert 0\rangle_{I int}$. 
	\item Region $II_{int}$ is the interior spatial region of SBH $r\lesssim 2GM$ with $U<0,\, V<0$ i.e., remaining half of the interior. The vacuum in this region is $\vert 0\rangle_{II int}$. 
	\item Region $I_{ext}$ is the exterior spatial region of SBH $r\gtrsim 2GM$ with $U<0,\, V>0$ i.e., half of the exterior. The vacuum in this region is $\vert 0\rangle_{I ext}$. 
	\item Region $II_{ext}$ is the exterior spatial region of SBH $r\gtrsim 2GM$ with $U>0,\, V<0$ i.e., remaining half of the exterior. The vacuum in this spatial region is $\vert 0\rangle_{II ext}$.
\end{itemize}
All this means is in the exterior, i.e., $r\gtrsim 2GM$, the quantum state 
\begin{equation}
	\begin{aligned}
		\hat{\Phi}_{ext} \vert 0\rangle_{ext} & = \frac{1}{\sqrt{2}} \LF \hat{\Phi}_{I ext}\oplus \hat{	\Phi}_{II ext}\RF \Big( \vert 0\rangle_{I ext} \oplus \vert 0\rangle_{II ext} \Big) \\ 
		& = \frac{1}{\sqrt{2}} \LF \hat{\Phi}_{I ext} \vert 0\rangle_{I ext}\oplus \: \hat{	\Phi}_{II ext} \vert 0\rangle_{II ext}\RF 
	\end{aligned}
	\label{extqf}
\end{equation}
is described by two components $\hat{\Phi}_{I ext} \vert 0\rangle_{I ext}$ and $\hat{\Phi}_{II ext} \vert 0\rangle_{II ext}$. 
Similarly, in the interior, i.e, $r\lesssim 2GM$, the quantum state is described by two components
\begin{equation}
	\begin{aligned}
		\hat{\Phi}_{int} \vert 0\rangle_{ext} & = \frac{1}{\sqrt{2}} \LF \hat{\Phi}_{I int}\oplus \hat{	\Phi}_{II int}\RF \Big( \vert 0\rangle_{I int} \oplus \vert 0\rangle_{II int} \Big)
		\\ 
		& = \frac{1}{\sqrt{2}} \LF \hat{\Phi}_{I int} \vert 0\rangle_{I int}\oplus \: \hat{	\Phi}_{II int} \vert 0\rangle_{II int}\RF 
	\end{aligned}
	\label{intqf}
\end{equation}
We represent these quantum fields in the conformal diagram Fig.~\ref{fig:fig4v3}, which illustrates the SBH DQFT, very similar to how we represented DQFT of Minkowski spacetime in Fig.~\ref{fig:minkowski-np}.  

{From \eqref{infcomm} and \eqref{Bhcomre}, we can easily deduce the relation between the interior and exterior creation and annihilation operators as
\begin{equation}
	\begin{aligned}
	\hat c_{I\, k} & = -i \frac{8\pi G}{r_S^2\LF \ell^2+\ell+1 \RF}	\hat b^\dagger_{I\, k} ,\quad 	\hat c_{II \,k} = -i \frac{8\pi G}{r_S^2\LF \ell^2+\ell+1 \RF}	\hat b^\dagger_{II \,k}  \\ 
		\hat c^\dagger_{I\, k} & = i \frac{8\pi G}{r_S^2\LF \ell^2+\ell+1 \RF}	\hat b_{I\, k} ,\quad 	\hat c^\dagger_{II \,k} = i \frac{8\pi G}{r_S^2\LF \ell^2+\ell+1 \RF}	\hat b_{II \,k} 
	\end{aligned}
	\label{intextop}
\end{equation}
The above relations establish the following connection between the vacuums of $I_{ext}$ and $I_{int}$ regions
\begin{equation}
	\begin{aligned}
	\hat c_{I\,k} \vert 0\rangle_{I int} & = -i \frac{8\pi G}{r_S^2\LF \ell^2+\ell+1 \RF}	\hat b^\dagger_{I\, k} \vert 0\rangle_{I int}  =0  \\ 
		\hat c^\dagger_{I\,k} \vert 0\rangle_{I ext} & = i  \frac{8\pi G}{r_S^2\LF \ell^2+\ell+1 \RF}	\hat b_{I\, k} \vert 0\rangle_{I ext}  =0
	\end{aligned}
	\label{vacintext}
\end{equation}
Similar relations can be derived for $II_{ext}$ and $II_{int}$ regions. The physical meaning of \eqref{vacintext} is as follows.
The creation operator of the exterior ($r\gtrsim 2GM$) vacuum acts as the annihilation operator of the interior ($r\lesssim 2GM$) vacuum. This would imply interior and exterior components of a single quantum state correspond to positive and negative energies as first envisioned by Hawking \cite{Hawking:1976ra}, and here we obtained with DQFT, which leads to unitarity\footnote{which will be discussed in the next section.}. This is due to the fact that the states in the interior and exterior SBH spacetime interact gravitationally.  
}

{This presents a new picture of understanding how the geometric superselection rule in our theory gets transformed in a non-trivial way as the following
\begin{equation}
	\begin{aligned}
	\LF \hat{c}_{I\,k}\oplus   \hat{b}_{I\,k}\RF \LF \vert 0\rangle_{I\,int}\oplus \vert 0\rangle_{I\,ext}\RF &  = \begin{pmatrix}
		\hat{c}_{I\,k} & 0 \\
			0 & \hat{b}_{I\,k}
	\end{pmatrix} \begin{pmatrix}
	\vert 0\rangle_{int} \\ \vert 0\rangle_{ext}
	\end{pmatrix} \\ & =  \begin{pmatrix}
	-i\frac{8\pi G}{r_S^2\LF \ell^2+\ell+1 \RF} \hat{b}^\dagger_{I\,k} & 0 \\
	0 & \hat{b}_{int} 
	\end{pmatrix} \begin{pmatrix}
	\vert 0\rangle_{int} \\ \vert 0\rangle_{ext}
	\end{pmatrix} = \begin{pmatrix}
	0 \\ 0
	\end{pmatrix}
	\end{aligned} 
\end{equation}
This means the exterior $I_{ext}$ region is a vacuum with respect to $b_{I,\,k}$ whereas the interior $I_{int}$ region is a vacuum with respect to $b^\dagger_{I\,k}$.  The relations \eqref{intextop} in \eqref{BHdecom} would imply 
\begin{equation}
	\begin{aligned}
	\hat \Phi^I_{int} &  =  \int \frac{dk}{\sqrt{2\pi}} \frac{1}{\sqrt{2\vert k\vert }} \Bigg[ \hat{c}_{I k} e^{i \vert k\vert T-i kX} + \hat{c}_{I k}^\dagger e^{-i \vert k\vert T+ikX}  \Bigg] \\
& =	i\frac{8\pi G}{r_S^2\LF \ell^2+\ell+1\RF}\int \frac{dk}{\sqrt{2\pi}} \frac{1}{\sqrt{2\vert k\vert }} \Bigg[- \hat{b}^\dagger_{I k} e^{i \vert k\vert T-i kX} + \hat{b}_{I k} e^{-i \vert k\vert T+ikX}  \Bigg] \\ & \neq  	i\frac{8\pi G}{r_S^2\LF \ell^2+\ell+1\RF}\hat{\Phi}^I_{ext}, 
	\end{aligned}
 \label{cbdagger}
\end{equation}
Combining \eqref{intextop} with the canonical relations \eqref{infcomm}, we obtain
\begin{equation}
	\begin{aligned}
		\Big[ \hat{\Phi}_{I int},\, \hat{\Phi}_{II ext}  \Big]  = 0 \\ 
		\Big[  \hat{\Phi}_{II int},\,\hat{\Phi}_{I ext}  \Big]  = 0\,. 
	\end{aligned}
	\label{extcom2}
\end{equation}
}
For an observer at $r\to \infty$ ($r_\ast \to \infty$) of SBH spacetime \eqref{SBHmet}, the quantum field is in Minkowski spacetime, and according to the DQFT formalism, we have
\begin{equation}
	\hat{\Phi}_{ext} = \frac{1}{\sqrt{2}}\LF \hat{\Phi}_{I \infty} \oplus \hat{\Phi}_{II \infty} \RF\,, 
 \label{infphi}
\end{equation}
where 
\begin{equation}
	\begin{aligned}
\hat{\Phi}_{I \infty}  & = \int_{0}^{\infty} \frac{d\tilde{\omega}}{\sqrt{2\pi}} \frac{1}{\sqrt{2\tilde{\omega} }} \Bigg[ \hat{a}_{I \tilde{\omega}} e^{-i\tilde{\omega} u} +  \hat{a}_{I -\tilde{\omega}} e^{-i\tilde{\omega} v} +  \hat{a}_{I \tilde{\omega}}^\dagger e^{i\tilde{\omega} u} + \hat{a}_{I -\tilde{\omega}}^\dagger e^{i\tilde{\omega} v}  \Bigg]  \\
\hat{\Phi}_{II \infty} &  = \int_{0}^{\infty} \frac{d\tilde{\omega}}{\sqrt{2\pi}} \frac{1}{\sqrt{2\tilde{\omega} }} \Bigg[ \hat{a}_{II -\tilde{\omega}} e^{i\tilde{\omega} u} +  \hat{a}_{II -\tilde{\omega}} e^{i\Omega v} +  \hat{a}_{II \tilde{\omega}}^\dagger e^{-i\tilde{\omega} u} + \hat{a}_{II \tilde{\omega}}^\dagger e^{-i\tilde{\omega} v}  \Bigg] ,
\end{aligned}
\end{equation}
where the creation and annihilation operators satisfy 
\begin{equation}
	\begin{aligned}
			\Big[\hat{a}_{I\,\tilde{\omega}},\,\hat{a}_{I\,\tilde{\omega}^\prime}^\dagger\Big] & = 		
			\Big[\hat{a}_{II\,\tilde{\omega}},\,\hat{a}_{II\,\tilde{\omega}^\prime}^\dagger\Big] = \hbar \delta^{(3)}\LF \tilde{\omega}-\tilde{\omega}^\prime \RF \\
				\Big[\hat{a}_{I\,\tilde{\omega}},\,\hat{a}_{II\,\tilde{\omega}^\prime}\Big] &=	\Big[\hat{a}_{I\,\tilde{\omega}},\,\hat{a}_{II\,\tilde{\omega}^\prime}^\dagger\Big] = \Big[\hat{a}_{I\,\tilde{\omega}}^\dagger,\,\hat{a}_{II\,\tilde{\omega}^\prime}^\dagger\Big]  =0\,.
			\end{aligned}
\end{equation}
This asymptotic Minkowski vacuum is given by
\begin{equation}
	\vert 0\rangle_{\infty} = \vert 0\rangle_{I \infty} \oplus \vert 0\rangle_{II \infty}, 
\end{equation}
where
\begin{equation}
\hat{a}_{I \tilde{\omega}}  \vert 0\rangle_{I\infty} =0,\quad \hat{a}_{II\, \tilde{\omega}} \vert 0\rangle_{II \infty} =0 \,. 
\end{equation}

By applying the technique of Bogoliubov transformation, we obtain the number density of particles created by operators $\hat{a}^\dagger_{I\,\omega},\, \hat{a}^\dagger_{II\,\omega}$ in the vacuum $\vert 0\rangle_{BH}$ which is what exactly asymptotic observer see as Hawking particles. Following \cite{Mukhanov:2007zz} we can compute
\begin{equation}
	\begin{aligned}
	a_{I \tilde{\omega}}  & = \int_{0}^{\infty} d\omega \LF \alpha_{\omega\tilde{\omega}} b_{I\,\omega}  +\beta_{\omega\tilde{\omega}} b^\dagger_{I\,\omega} \RF \\
		a_{II \tilde{\omega}}  & = \int_{0}^{\infty} d\omega \LF \tilde{\alpha}_{\omega\tilde{\omega}} b_{II\,\omega}  +\tilde{\beta}_{\omega\tilde{\omega}} b^\dagger_{II\,\omega} \RF
	\end{aligned}
\end{equation}
where
\begin{equation}
	\begin{aligned}
		\alpha_{\omega\tilde{\omega}} & = \sqrt{\frac{\tilde{\omega}}{\omega}} \int_0^\infty  \frac{du}{2\pi} e^{i\tilde{\omega} u-i\tilde{\omega} U} \quad 	\beta_{\omega\tilde{\omega}} & = \sqrt{\frac{\tilde{\omega}}{\omega}}\int_0^\infty  \frac{du}{2\pi} e^{i\tilde{\omega} u+i\tilde{\omega} U} \\ 
		\tilde{	\alpha}_{\omega\tilde{\omega}} & = \sqrt{\frac{\tilde{\omega}}{\omega}}\int_0^\infty  \frac{du}{2\pi} e^{-i\tilde{\omega} u+i\tilde{\omega} U} \quad 	\tilde{	\beta}_{\omega\tilde{\omega}} & = \sqrt{\frac{\tilde{\omega}}{\omega}}\int_0^\infty  \frac{du}{2\pi} e^{-i\tilde{\omega} u-i\tilde{\omega} U}
	\end{aligned}
\end{equation}
Given all the above relations, the asymptotic observer sees the number density 
\begin{equation}
	\begin{aligned}
 N_{\tilde{\omega}} & = \frac{1}{2}	{}_{BH}\langle 0 \vert \LF \hat{a}_{I\,\tilde{\omega}}^\dagger \oplus \hat{a}_{II\,\tilde{\omega}}^\dagger  \RF \LF \hat{a}_{I\,\tilde{\omega}} \oplus \hat{a}_{II\,\tilde{\omega}} \RF  \vert 0\rangle_{BH}  \\ 
& = \frac{1}{2} \int_{0}^{\infty}d\omega \LF \vert \beta_{\omega\tilde{\omega}} \vert^2+ \vert \tilde{	\beta}_{\omega\tilde{\omega}} \vert^2 \RF\, \\ 
& = \frac{1}{e^{8\pi GM\tilde \omega/\hbar}-1}\,,
\end{aligned}
\label{numberBH}
\end{equation} 
where the prefactor $\frac{1}{2}$ is because of the prefactor in \eqref{infphi}. {It is worth noting that the result in \eqref{numberBH} remains unchanged when derived using the Bogoliubov transformation between the vacuum states of Minkowski spacetime (representing an infinitely long time before the black hole formation) and Kruskal-Szekeres (KS) spacetime (representing an infinitely long time after the black hole formation) \cite{Mukhanov:2007zz}, as originally outlined in Hawking's seminal work \cite{Hawking:1975vcx}. This consistency is expected, as Hawking radiation fundamentally arises from the Bogoliubov transformation between the vacuum states in the KS metric of a Schwarzschild black hole \eqref{KSmet} and the asymptotic Minkowski metric at {($r\to \infty)$}. }

Applying thermodynamics, the Hawking temperature of the BH is (in the units of $G=1$)
\begin{equation}
	T = \frac{\hbar}{8\pi M}
\end{equation}
Note that once we derive the number density of the particles that a distant observer can see, we can apply (classical) thermodynamics to understand how BH evaporates following the footsteps of Hawking and Bekenstein \cite{Hawking:1975vcx,Bekenstein:1973ur}. Following the Stefan-Boltzmann law, the mass of the BH  decreases as 
\begin{equation}
	\frac{dM}{dt} =  -\frac{{\Gamma} n_\gamma}{15360\pi M^2}\,,
 \label{timedec}
\end{equation}
where $n_\gamma$ denotes the number of degrees of freedom {and $\Gamma$ is a coefficient associated with the greybody factors \cite{Mukhanov:2007zz}.}
Applying the first law of thermodynamics, the entropy of the BH is estimated to be \cite{Mukhanov:2007zz}
\begin{equation}
	S_{BH} = 4\pi M^2\,. 
	\label{entropy}
\end{equation}
In the next sections, we show that the Hawking radiation we derived here with DQFT gives a thermal spectrum with pure states. This is what is needed to achieve unitarity and observer complementarity. In a recent paper, we have studied DQFT in Rindler spacetime \cite{Kumar:2024oxf}, and we have shown that one can achieve a thermal spectrum for every observer in Rindler spacetime preserving unitarity. Some of these details are briefly presented in Appendix.~\ref{sec:Rindler}. It is especially interesting to notice similarities between Fig.~\ref{fig:RindlerST} and \ref{fig:fig4v3}, which elucidates universal features of DQFT in different spacetimes with horizons.

\subsubsection{Unitarity and BH DQFT solution to the information paradox} 

\label{sec:unitarity}

From the beginning of understanding (quantum) BHs \cite{Hawking:1975vcx,Hawking:1976ra}, it has been assumed that the Hawking radiation is a pair production of particles in the interior (very near to the horizon $r\lesssim 2GM$) and the exterior (very near to the horizon $r\gtrsim 2GM$) and their combined Hilbert space is the direct product of individual Hilbert spaces \cite{Kim:2022pfp,Mathur:2009hf,Calmet:2022swf} 
\begin{equation}
	\Hc_{P} = \Hc_{int}\otimes \Hc_{ext}
	\label{entangle}
\end{equation}
This originates an external particle that a distant observer sees in a mixed state. This has been projected to originate the unitarity violation, given the assumption of BH formation with pure states \cite{Page:1993wv,Almheiri:2020cfm,Maldacena:2018izk}. Furthermore, it was shown by Don Page that the entanglement entropy between the particle pair grows and can exceed the entropy of the SBH when half of the BH mass has evaporated \cite{Page:1993wv,Page:2013dx}. To save unitarity, Don Page proposed that the entanglement entropy of the Hawking radiation must start to decrease after some time (Page time) so that the unitarity is recovered after the complete evaporation of BH \cite{Page:1993wv,Page:2013dx}. This is called the Page curve \cite{Almheiri:2012rt} and to obtain Page curve, several quantum gravity proposals have emerged in recent years \cite{Almheiri:2020cfm,Raju:2020smc}. 

From our point of view, the failure of unitarity, in the sense of Hawking radiation, is a mixed state, and the growth of entanglement entropy completely relies on the Hilbert space structure of Hawking pair \eqref{entangle}. The BH DQFT proposal leads to the direct-sum-Fock space \eqref{fockBH} 
and it leads to the Hilbert space of the Hawking pair as 
\begin{equation}
	\begin{aligned}
	\Hc_{\Pc} & =  \Hc_{int} \oplus \Hc_{ext}\,, 
	\end{aligned}
	\label{no-entangle}
\end{equation}
This leads to a pure (maximally) entangled state that becomes direct-sum of two pure states of the geometric superselection sectors $\LF  \Hc_{int},\, \Hc_{ext} \RF$ as
\begin{equation}
	\begin{aligned}
    \vert \Tilde{\psi}_{12}\rangle =\sum_{m,n} \Tilde{c}_{mn} \vert \Tilde{\phi}_{m1}\rangle  \otimes \vert \Tilde{\phi}_{n2}\rangle & = \frac{1}{2}\sum_{m,n} \Tilde{c}_{mn} \LF \vert \Tilde{\phi}_{m1}^{ext}\rangle \oplus \vert \Tilde{\phi}_{m1}^{int}\rangle\RF   \otimes   \LF \vert\Tilde{\phi}_{n2}^{ext}\rangle \oplus \vert \Tilde{\phi}_{n2}^{int}\rangle\RF 
    \\& = \frac{1}{2}\sum_{m,n} \Tilde{c}_{mn} \LF \vert \Tilde{\phi}^{\rm ext}_{m1}\rangle  \otimes \vert \Tilde{\phi}^{\rm ext}_{n2}\rangle \RF \oplus \LF \vert \Tilde{\phi}^{\rm int}_{m1}\rangle  \otimes \vert \Tilde{\phi}^{\rm int}_{n2}\rangle \RF \\ & = \vert \tilde{\psi}_{12}^{ext}\rangle  \oplus \vert \tilde{\psi}_{12}^{int}\rangle 
    \end{aligned}
\end{equation}
where $\Tilde{c}_{mn}\neq \Tilde{c}_n\Tilde{c}_m$, $\vert \phi_1\rangle = \sum_m \tilde{c}_m\vert \phi_{m1}\rangle  $ and $\vert \phi_2\rangle = \sum_n \tilde{c}_n\vert \phi_{n2}\rangle  $, 
the superscripts "ext" mean field component for $r\gtrsim 2GM$ whereas "int" mean field component for $r\lesssim 2GM$.
{This implies Hawking's pair of quanta $\vert \tilde{\psi}_{12}\rangle$ (pure state)  is divided into two pure state components $\vert \tilde{\psi}^{ext}_{12}\rangle$ and $\vert \tilde{\psi}^{int}_{12}\rangle$ of the geometric superselection sectors. Thus, the density matrix ($\rho_{H}$) of Hawking (pair) quanta becomes direct-sum of exterior and interior components 
\begin{equation}
    \rho_{H} = \rho_{ext}\oplus \rho_{int}
\end{equation}
Since exterior and interior are geometric superselection sectors, the Von Neumann entropies of exterior and interior (pure) state components ($S_{ext},\, S_{int}$) vanish
\begin{equation}
    S_{ext} = -Tr\LF\rho_{ext}\ln \rho_{ext}\RF=0,\quad S_{int} = -Tr\LF\rho_{int}\ln \rho_{int}\RF=0,\quad S_{H} = S_{ext}+S_{int}=0\,,
\end{equation}
which means every observer witnesses pure states, which results in black hole complementarity. Note that the way we achieve this is different from what was proposed in \cite{Almheiri:2012rt} because our approach has no firewalls\footnote{Since we assume the local Minkowski vacuum for an infalling observer.}, Hawking radiation is in a pure state, and there is observer complementarity, and everything is done within GR and quantum mechanics. In our picture, a single quantum state is spread across the interior and exterior as a direct-sum of two components, which is different from the standard quantization applied by Hawking, which assumes the notion of time (and its arrow) is the same everywhere.} 

The rule of direct-sum separates the Hilbert spaces into two separate spatial regions (superselection sectors), and we cannot have a superposition of states between those regions. This is very similar to the superselection rule proposed by Wick, Wightman, and Wigner \cite{Wick:1952nb}. Due to the Hilbert space structure \eqref{no-entangle}, we do not have any Page curve and no unitarity violation. We get unitary physics for quantum states in the interior and exterior. A similar thing happens when one does DQFT in Rindler spacetime \cite{Kumar:2024oxf}, which we briefly discuss in Appendix.~\ref{sec:Rindler}.  

Coming to the information problem question basically arises from the fact that the creation and annihilation operators of the interior and exterior field operators are taken to commute (See Eq. (2.8)-(2.10) of \cite{Hawking:1976ra}). But assuming 't Hooft's quantum commutation relations, derived from gravitational backreaction \eqref{noncompm}, we arrived at \eqref{intextf} which relates interior and exterior fields in a very non-trivial way. This is the first step to solving the information problem. 
This means that by observing Hawking radiation from outside, we should be able to make computations to probe the interior. This subject needs further investigation of BH DQFT and we leave it for future investigations. 

\subsubsection{Black hole is not an ordinary quantum system}

According to the so-called central dogma \cite{Almheiri:2012rt}, BH must be an ordinary quantum system. Actually, this means the Hilbert space of the degrees of freedom of the subsystem and the environment must be entangled by direct-product operation \eqref{entangle}. This means the concept of time should be the same in the subsystem and the environment. This is not the situation for BH, and we argue that BH cannot be an ordinary quantum system, and our proposal goes against the central dogma. 

If the central dogma is not true, it was conjectured that we should see so-called firewalls \cite{Almheiri:2012rt} i.e., a freely falling observer should see particles that violate the equivalence principle. We ascertain that this is not our case because, in the near horizon limit, \eqref{nearHlim} vacuum \eqref{BHvac} is locally Minkowski as explained after \eqref{vacBHop}.
Therefore, BH DQFT leads to unitary quantum physics without firewalls and without any Page curve and Page time.

\section{Questions of (quantum) black hole formation and evaporation}
\label{sec:evapBH}

The idea of this section is to make a logical guess on what a conformal diagram of BH DQFT can look like. Classically, the description of the BH spacetime diagram formed by collapse looks like Fig.~\ref{fig:collapsingbh} as Hawking initially described \cite{Hawking:1975vcx}. However, it is very unclear how to give a quantum mechanical (microscopic) description of collapse. During the collapse, we have matter imploding, and the horizon of the BH (i.e., the time-dependent Schwarzschild radius) grows with time, which is expected to be a much more rapid process than evaporation when the horizon size (the time-dependent Schwarzschild radius) starts to shrink. Of course, for the high energy modes, collapse can be described as an ultra-high energy superplanckian scattering \cite{Giddings:2001bu,Chen:2021dsw,Gaddam:2020mwe,Gaddam:2020rxb,Sanchez:2018lnb,Sanchez:2020rqj,Sanchez:2023ylf}.  It is still an open question to understand the transition between Minkowski spacetime QFT and the context when gravitational effects on quantum fields cannot be neglected. To be precise, the scattering of particles under the effect of dynamical spacetime is the key issue to be understood. Collapse or evaporation of a (quantum) BH means that, macroscopically, there is a direction of the system time evolution encoded in the growing or shrinking of the dynamical horizon.  The macroscopic dynamical effects will break $\Pc\Tc$ symmetry of the metric, which would affect the evolution of microscopic degrees of freedom. {A violation of time reversal symmetry happens for an evaporating SBH or during the gravitational collapse. This
should distort the conformal diagram of SBH Fig.~\ref{fig:fig4v3}, where the regions with the different components of field that evolve backward in time $\hat \Phi_{II\,ext}\vert \hat 0_{II\,ext}\rangle$ behave differently from the components of field that evolve forward in time $\hat \Phi_{I\,ext}\vert \hat 0_{I\,ext}\rangle$. This is because quantum fields in DQFT always have evolution with opposite arrows of time at parity conjugate points. If the classical spacetime breaks the time-reversal symmetry, the quantum fields respond accordingly in DQFT. }
 This is similar to the recent study of how quantum fields in an expanding Universe would create parity asymmetry in cosmic microwave background \cite{Gaztanaga:2024vtr,Gaztanaga:2024whs,Kumar:2022zff,GKM}. We anticipate a similar effect for BH physics, but we seriously lack any machinery to do such computations. Therefore, we leave this for future investigations.

\section{BH DQFT implications for S-matrix and quantum gravity}
\label{sec:QHQG}

S-matrix is the most important notion in QFT, and it is supposed to describe the scattering of particles. The notion of S-matrix has varied meanings in curved spacetime, and most often, it is not considered as a required concept. However, some questions remain, such as: how do the standard model particles we know in nature react to curved spacetimes with horizons? How does a particle, and if it decays, get affected by the presence of curved spacetime? Therefore, if not the S-matrix, we need something similar to describe particle scatterings. There is another notion of S-matrix that has to do with the complete formation and full evaporation of BH, which definitely requires a full understanding of quantum gravity and currently, we do not have any definite candidate for that. Given we have multiple frameworks of quantum gravity \cite{Bousso:2022ntt,Harlow:2022qsq} and currently we essentially need quantitative ideas to test quantum gravity with astrophysical and gravitational wave observations to anchor what direction is correct for quantum gravity. Even to make significant advancements in quantum gravity research, we do need to address the issue of quantum fields in curved spacetime. 

 BH DQFT quantization gives an elegant way to separate the BH interior from the exterior without violating unitarity and, at the same time, thanks to the 't Hooft quantum backreaction algebra, we do know now how interior and exterior degree of freedom are connected with each other non-trivially \eqref{intextf}. With this, we can aim to understand first particle interactions, scattering, and decays near the horizon. To an exterior observer, all the particle interactions expected happen unitarily, of course, if degrees of freedom disappear in the interior of the BH they would definitely leave an imprint in the behavior of the degrees of freedom outside by \eqref{intextf} or its modification when we include interactions. 

\section{Conclusions and outlook}
\label{sec:Conc}

In this section, we aim to give a quick summary of what was done in this paper. Since the proposal of Hawking's seminal paper on BHs, we have been highly grappled with issues related to information paradox, unitarity violation, and breakdown of predictability. All of these emerged from quantizing a scalar field in SBH spacetime. The identification of this problem even dates back to 1935, when Einstein and Rosen pointed out the "Particle problem in General Relativity" \cite{Einstein:1935tc} (and also see \cite{GKM}), which exposes the fundamental problem between gravity and quantum mechanics. 
Several developments have happened in the last decades under the scope of quantum gravity to understand BHs and the several enigmas associated with them. Don Page's calculation of the entanglement entropy growth beyond the BH entropy has bolstered even more the puzzles associated with BHs. In this paper, we paid very critical attention to the origin of inconsistencies in our understanding of BHs with quantum mechanics. We noticed the importance of \cpt\, in the standard QFT and the questions related to the status of \cpt\, in BHs physics which were consistently asked by Hawking and 't Hooft seminal works in the last decades. This has caught our attention in building QFT with special emphasis on discrete spacetime transformations, which are vital in quantum theory. Acknowledging the fact that time is a parameter in quantum theory, in this paper, we formulated a direct-sum QFT with geometric superselection rules in the context of BH spacetime based on our previous study in the context of Minkowski, de Sitter, and inflationary spacetimes\footnote{The framework of DQFT for inflationary quantum fluctuations is found to explain long-standing CMB anomalies \cite{Gaztanaga:2024vtr} 650 times better than the standard theory of inflation. Thus, our framework of DQFT passes the needed observational test and can potentially lead to new unknown effects in BH physics. } \cite{Kumar:2023ctp,Gaztanaga:2024vtr}. We showed that our DQFT quantizations does preserve the unitarity and observer complimentarity in  a non-trivial way which is different from the earlier proposals of AMPS \cite{Almheiri:2012rt,Almheiri:2020cfm}.

Our prescription of quantization, together with 't Hooft gravitational backreaction effects, has revealed a new structure for understanding quantum fields in BH spacetime. Note that all our study is restricted to Schwarzschild BH in this paper. Our BH DQFT implies no entanglement between interior and exterior Hawking particles pairs, at the same time we obtained that the exterior radiation is not independent of the interior due to the new set of non-commutative relations originated from 't Hooft gravitational backreaction effects. 
This effect is absent from Hawking's original 1975 paper. The highlight of our construction is that Hawking radiation is in a pure state, with no firewalls at the BH horizon, and we achieve observer complementarity. 
We have discussed in detail the implications of our result for further explorations in treating quantum fields in collapse and evaporation of BH and future research in quantum gravity. 

\acknowledgments

KSK was supported by JSPS and KAKENHI Grant-in-Aid for Scientific Research No. JP20F20320 and No. JP21H00069. KSK would like to thank the Royal Society for supporting in the name of Newton International Fellowship. 
KSK would like to thank Mainz U. and Beira Interior U. for their hospitality, which is where part of the work has been carried out. 
J. Marto is supported by the grant UIDB/MAT/00212/2020 and COST Action CA23130. We want to thank Chris Ripken for the initial collaboration on the project and for giving us a lot of insights into the algebraic QFT.  We thank Gerard 't Hooft for truly inspiring us over the years with his impactful insights into the subject. Also, we thank him for the very useful comments and discussions while preparing this manuscript. We thank Yashar Akrami, Norma G. Sanchez, Masahide Yamaguchi, Alexei A. Starobinsky, Luca Buoninfante, Francesco Di Filippo, Yasha Neiman, Paolo Gondolo, Martin Reuter, Paulo V. Moniz, Gia D'vali, Mathew Hull, Enrique Gazta\~naga for useful discussions. KSK would like to especially thank Sivasudhan Ratnachalam for his positive encouragement, friendship, and useful discussions on quantum mechanics. 

\appendix 

\section{DQFT in Rindler spacetime}

\label{sec:Rindler}

In this section, we briefly review the formulation of DQFT  in Rindler spacetime, which worked out in \cite{Kumar:2024oxf}. This section will give a complementary understanding of DQFT in BH spacetime with the DQFT in Rindler spacetime Sec.~\ref{sec:NotPen}.

\begin{figure}
    \centering
    \includegraphics[width=0.5\linewidth]{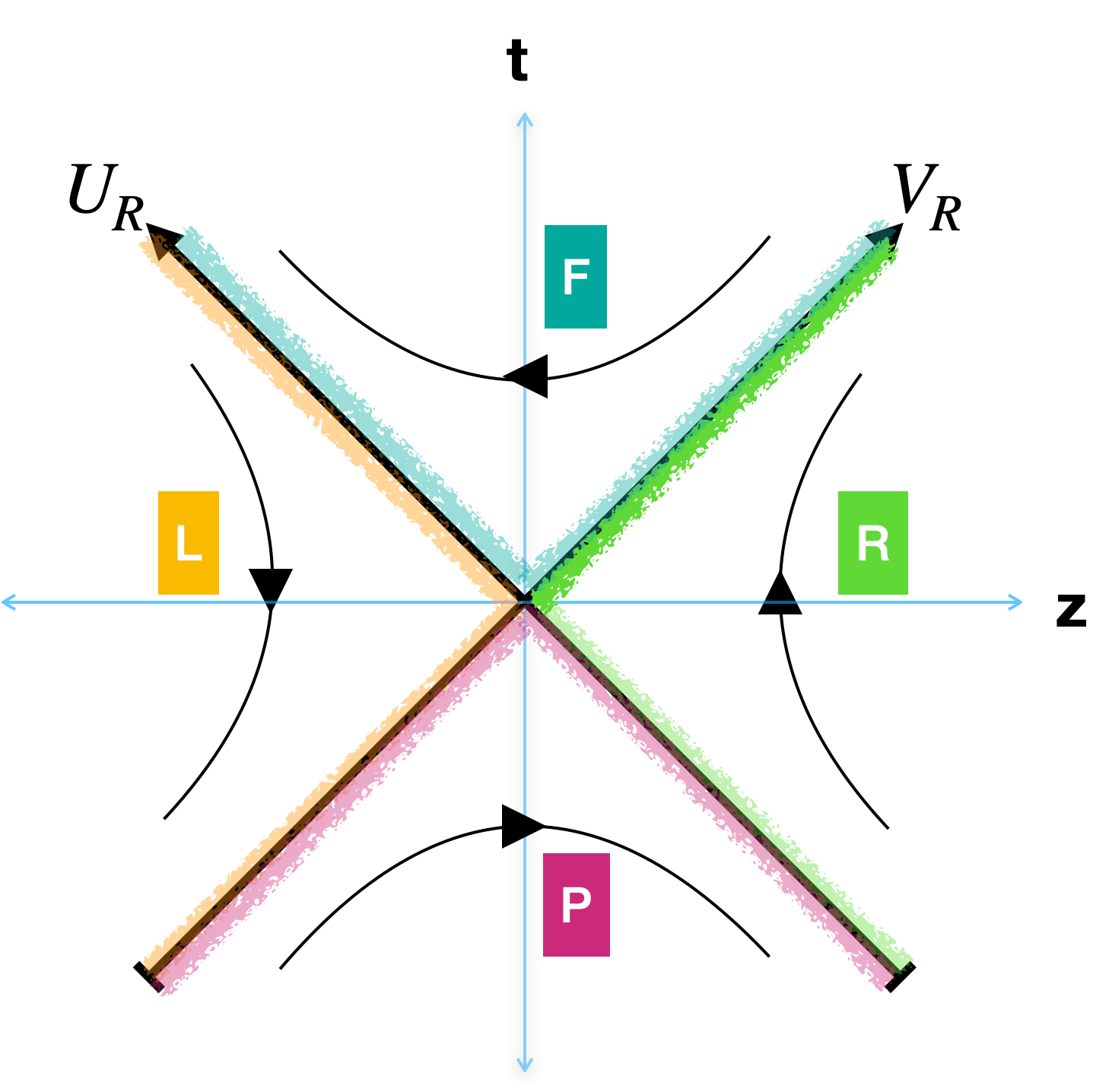}
    \caption{In this picture, we depict the Rindler spacetime with the Left, Right ($z^2\gtrsim t^2$), and Future, Past ($t^2\gtrsim z^2$) regions in Minkowski plane $\LF t,\,z \RF$. The curved lines with arrows in the Left and Right regions depict a constant acceleration $ae^{-a\xi}$ in the direction $\eta: \infty \to -\infty$ and $\eta: -\infty \to \infty$ respectively. Future and Past Rindler wedges are the degenerate Kasner Universes where the arrows are changing $z: \pm\infty \to \mp\infty$, which means $\eta: \mp\infty \to \pm\infty$. The fuzzy colored lines indicate Rindler Horizons for Left (Yellow), Right (Green), Future (Cyan), and Past (Pink).}
    \label{fig:RindlerST}
\end{figure}

Rindler spacetime is the coordinate transformation of Minkowski spacetime. Here, we consider (1+1) dimension Minkowski spacetime represented by
\begin{equation}
    ds^2 = -dt^2+ dz^2
\end{equation}
\begin{equation}
\begin{aligned}
 z^2-t^2 & = \frac{1}{a^2} e^{2a\xi} \implies \begin{cases}
     z  = \frac{1}{a} e^{a\xi} \cosh{a\eta},\quad  t= \frac{1}{a} e^{a\xi} \sinh{a\eta} \quad \LF \textrm{Right Rindler} \RF \\   z  = -\frac{1}{a} e^{a\xi} \cosh{a\eta},\quad  t= \frac{1}{a} e^{a\xi}\sinh{a\eta} \quad \LF \textrm{Left Rindler} \RF
 \end{cases} \\ &\implies \boxed{ds^2 = e^{2a\xi}\LF -d\eta^2+d\xi^2\RF} \\   
  t^2-z^2 & = \frac{1}{a^2} e^{2a\eta} \implies \begin{cases}
     t  = \frac{1}{a} e^{a\eta} \cosh{a\xi},\quad  z= \frac{1}{a} e^{a\eta} \sinh{a\xi}\quad \LF \textrm{Future Kasner} \RF  \\  t  = -\frac{1}{a} e^{a\eta} \cosh{a\xi},\quad  z= \frac{1}{a} e^{a\eta} \sinh{a\xi}\quad \LF \textrm{Past Kasner} \RF 
 \end{cases} \\ & \implies \boxed{ds^2 = e^{2a\eta}\LF -d\eta^2+d\xi^2\RF}
    \end{aligned}
    \label{Robserco}
\end{equation}
The description of this spacetime is presented pictorially in Fig.~\ref{fig:RindlerST}.

The Rindler coordinates can be rewritten as 
\begin{equation}
    \begin{aligned}
        U_R &= -\frac{1}{a}e^{-au}<0,\quad &&V_R= \frac{1}{a}e^{av}>0\quad &&(\rm Right\, Rindler)\\
        U_R&= \frac{1}{a}e^{-au}>0,\quad &&V_R= -\frac{1}{a}e^{av}<0\quad &&(\rm Left\, Rindler) \\
         U_R&= \frac{1}{a}e^{-au}>0,\quad &&V_R= \frac{1}{a}e^{av}>0\quad &&(\rm Future\, Kasner) \\
          U_R&= -\frac{1}{a}e^{-au}<0,\quad &&V_R= -\frac{1}{a}e^{av}<0\quad &&(\rm Past\, Kasner)
    \end{aligned}
    \label{UVcoord}
\end{equation}
where 
\begin{equation}
\begin{aligned}
    u&= \eta-\xi,\quad v=\eta+\xi \\ 
    U_R&= t-z,\quad V_R=t+z
    \end{aligned}
\end{equation}
These coordinates \eqref{UVcoord} define the Left, Right, Future, and Past Rindler regions separated by horizons carried through discrete transformations on $\LF U_R,\, V_R \RF$. 
These are analogous to the Kruskal coordinates of SBH in \eqref{rg2M} and \eqref{rg2M2}. Comparing Figs.~\ref{fig:fig4v3} and \ref{fig:RindlerST}, we can draw similarities between DQFT in BH and Rindler spacetimes. 

According to DQFT in Rindler spacetime \cite{Kumar:2024oxf}, a (maximally) entangled pure state (for $z^2-t^2\gtrsim 0$) is a direct-sum of two pure state components in Left Rindler and Right Rindler regions 
\begin{equation}
    \vert \psi_{LR}\rangle = \frac{1}{\sqrt{2}}\begin{pmatrix}
        \vert \phi_{R1} \rangle \otimes \vert \phi_{R2} \rangle \\ 
        \vert \phi_{L1} \rangle \otimes \vert \phi_{L2} \rangle
    \end{pmatrix}
    \label{LRent}
\end{equation}
Similarly, one can also deduce the same for Future and Past regions. Since each component in \eqref{LRent} is a pure state, the Von Neumann entropy for each component vanishes; thus, both observers (Left and Right) perceive pure states evolving into pure states.

\bibliographystyle{utphys}
\bibliography{SBH.bib}

\end{document}